\pdfoutput=1
\documentclass[sigconf,nonacm,9pt]{acmart}

\AtBeginDocument{%
  \providecommand\BibTeX{{%
    \normalfont B\kern-0.5em{\scshape i\kern-0.25em b}\kern-0.8em\TeX}}}

\setcopyright{none}
\usepackage{pifont}
\usepackage{pdfpages}
\usepackage{circledsteps}
\usepackage{tabularray}
\usepackage{amsmath}
\usepackage{hhline}
\usepackage{booktabs}
\usepackage{tabularx}
\usepackage{graphicx}
\usepackage{float}
\usepackage{multirow}
\usepackage{tikz}
\usepackage{xcolor,colortbl}
\usepackage{titlesec}
\usepackage{enumitem}
\usepackage{xspace}
\usetikzlibrary{tikzmark,calc}

\newcommand{\sys}{\texttt{EventHOI}\xspace}

\usepackage{xcolor}

\newcommand{\presec}{\vspace{-0.03in}}

\begin{document}

\title{\sys: Event-Driven Resource-Efficient Hand-Object Interaction Logging on Smartglasses}

\author{Le~Zhang$^{1}$, Kailai~Cui$^{1}$, Hao~Chen$^{2}$, Vlad~Roznyatovskiy$^{2}$, Jianzhong~Zhang$^{2}$, Anhong~Guo$^{1}$, Kang~G.~Shin$^{1}$, Ke~Sun$^{1}$}
\affiliation{%
  \institution{$^{1}$University of Michigan\qquad $^{2}$Samsung Research America}
  \country{}}
\email{zle@umich.edu}
\renewcommand{\shortauthors}{Zhang et al.}

\begin{abstract}
Hand-object interactions (HOIs) can be captured by smartglasses to learn
how people manipulate objects in daily life.
Automatic logging of these interactions from an egocentric perspective provides
time-stamped evidence of daily actions for applications like memory assistance.
However, continuous RGB capture and wireless transmission exceed the energy and bandwidth budgets of smartglasses.
We present \sys, an event-driven sensing and compression system that couples 
selective RGB capture with HOI-aware encoding to enable egocentric logging over Bluetooth Low Energy (BLE) while preserving downstream HOI logging performance.
Our key insight is to use low-power event sensing both to determine \emph{when} RGB capture is needed and to identify \emph{where} visual detail must be preserved.
A lightweight IMU-based egomotion cancellation pipeline suppresses 
egomotion-induced events so that residual HOI motion can guide RGB triggering and spatial compression.
We build a smartglasses prototype and conduct an end-to-end evaluation of \sys on a 50-hour egocentric benchmark with 97 subjects and a 20-hour real-world dataset we collect from 10 participants, together covering a broad range of daily-life scenarios and environments.
Compared to RGB-only baselines, \sys reduces always-on backend power consumption 
by 5.8--10.4$\times$ and uplink bandwidth by 2.9$\times$, with minimal HOI 
recognition accuracy loss.
The resulting RGB uplink of less than 130~kbps fits within commodity BLE bandwidth budgets, making stable BLE video streaming feasible for egocentric HOI logging.
Demo video and project website: \url{https://eventhoi.github.io}.
\end{abstract}

\maketitle

\vspace{-0.15in}
\presec \section{Introduction}

\begin{figure}[t]
    \centering
    \includegraphics[width=0.8\columnwidth]{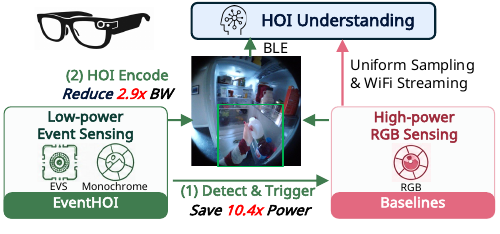}
    \vspace{-0.1in}
    \caption{\sys uses low-power event sensing to (1) detect and selectively trigger the high-power RGB camera and (2) compress RGB uplink for HOI understanding.}
    \label{fig:introduction}
    \vspace{-0.2in}
\end{figure}
Hand–Object Interaction (HOI) is an atomic unit of everyday activities.
Logging HOIs provides time-stamped evidence of daily actions, i.e., what
objects the hands interacted with, what actions occurred, and when, enabling
immediate assistance in daily life.
This capability is important because everyday memory slips are both
common and consequential: diary-based studies show that over 60\%
of reported memory failures involve forgetting intended
actions~\cite{mogle2023measuring}.
As a result, HOI logging can naturally support \textit{reminders}
(e.g., ``Did I lock the door?''), \textit{verification} (e.g., ``Did
I take my medication?''), and rapid \textit{retrieval} (e.g., ``Where
did I leave the charger?''), which correspond to frequent types
of day-to-day lapses.
Beyond memory augmentation, HOI logs can also strengthen health-routine
\textit{summarization}: the World Health Organization (WHO) reported
that adherence to long-term therapies for chronic diseases averages
only $\sim$50\% in developed countries, motivating systems that can
passively verify routine actions without relying on self-report~\cite{rassool2004failure}.

Recent camera-equipped smartglasses have increased interest in
egocentric HOI logging by providing a wearable, always-available
viewpoint that naturally captures the wearer’s hands and the objects
they manipulate.
This trend is reflected in the rapid growth of large-scale
egocentric HOI datasets and advances in computer vision (CV) algorithms
\cite{ma2024nymeria, grauman2022ego4d, perrett2025hd}.
Despite this momentum, a key bottleneck remains: \textit{prior
work often implicitly assumes away the on-device system cost
of continuous sensing on smartglasses.}
In practice, egocentric HOI logging demands \textit{(1)} continuous RGB
recording and \textit{(2)} sustained video
uplink to edge/cloud resources for CV processing.

For continuous RGB recording, Ray-Ban Meta glasses cap each
video clip at up to 3 minutes, due in part to overheating
during sustained use~\cite{metarayban_maxvideolength}.
Even research prototypes face tight energy budgets: Meta Aria
Gen~1 uses a 2.5~Wh battery and supports only $\sim$1.5 hours
of continuous local recording~\cite{aria_gen1_specs}.
Since larger batteries compromise ergonomics and user comfort of wearable devices, energy-efficient RGB sensing is essential to reduce power consumption and enable all-day recording.
Prior research such as modality-driven RGB triggering using thermal cameras or microphones incurs high energy overhead
or lacks robustness in noisy environments~\cite{paruchuri2025egotrigger,shahi2025thor},
while event camera only solutions focus on close-set gesture classification~\cite{bhattacharyya2024helios},
making real-world application and deployment infeasible.

Additionally, smartglasses rely on wireless connectivity to offload RGB video for cloud processing.
Wi-Fi video offloading is too energy-intensive for all-day operation on smartglasses~\cite{naderiparizi2017glimpse}. BLE offers a lower-power alternative, but its throughput can be as low as 0.3\,Mbps, below the bitrate required by our RGB-streaming baseline to preserve HOI recognition accuracy (\S\ref{sec:prelim-study-2})~\cite{infineon2020blethroughput}.
Traffic from microphones, IMUs, and other sensing modalities further reduces the available bandwidth.
Recent semantic video compression approaches
can reduce bitrate substantially,
but they typically rely on heavy on-device computation,
such as learning-based saliency estimation, making them difficult to deploy on resource-constrained smartglasses~
\cite{hojjat2024limitnet, xiao2024task, chen2026kdc}.
These techniques are also rarely evaluated on egocentric streams,
where camera motion and scene dynamics change continuously
with the wearer’s actions.

In this paper, we design \sys to reduce the on-device system cost
of egocentric HOI logging on smartglasses on sensing, computation, and communication.
Our key idea is to design \textit{an event-driven low-power sentinel}, using low-power event sensing to detect
HOI-induced motion and selectively control high-power RGB
camera recording and compression.
As shown in Fig.~\ref{fig:introduction}, we envision future
smartglasses equipped with a low-power modality, such as \textit{an
event camera} or \textit{a monochrome camera}, running continuously to
produce event frames.
A lightweight on-device pipeline processes these event frames as an always-on HOI detector that can \emph{(i)} localize HOI
timestamps and \emph{(ii)} estimate an HOI saliency region.
These cues then steer the RGB camera to record only when an HOI is
detected and enable saliency-guided spatial compression, so that
the compressed RGB stream can be uploaded directly to an edge/cloud
device over commodity Bluetooth Low Energy (BLE), without
requiring network-layer modifications.
The uploaded RGB clips are then analyzed by powerful
CV models, e.g., vision-language models (VLMs), on the edge/cloud
to produce accurate HOI logs.

\sys must address the following three key challenges.\\
\textit{\underline{Challenge 1}: How can we design a lightweight
algorithm to cancel egomotion in event frames?}
Egocentric event streams are often dominated by wearer-induced
motion (e.g., head turns and walking), which can easily overwhelm
the subtler dynamics introduced by HOIs.
To make always-on HOI detection practical, we leverage the on-board 6-axis IMU available on most smartglasses to drive a two-stage global
event cancellation pipeline.
First, a lightweight model estimates instantaneous egomotion speed
from IMU signals, avoiding drift from long-term integration, and
filters out large translational motion such as walking.
Second, \sys uses gyroscope measurements to estimate the
rotation-induced head-motion field and subtract the corresponding
global events from the event frame, mitigating interference from
head rotations.

\noindent\textit{\underline{Challenge 2}: How can we reliably detect
HOIs and estimate an HOI saliency region from the event frames
for selective RGB triggering and spatial compression?}
Even after egomotion cancellation, HOI signatures in event frames
remain sparse, short-lived, and diverse across actions, objects,
and environments.
Residual events may also include non-HOI motion, such as moving
bystanders or background changes, causing spurious triggers.
To address this, \sys designs an event-driven HOI detector that
exploits a simple egocentric prior: during HOIs, the wearer’s
hands and forearms occupy a sufficiently large and consistently
active region in the field of view.
Building on this, we model the spatiotemporal structure of
residual events and aggregate evidence over a short temporal
window to localize HOI timestamps robustly.
Given a detected HOI window, we estimate an HOI saliency region
by identifying consistently activated event clusters and refining
them into a compact, temporally stable region of interest (RoI),
enabling reliable saliency-guided spatial compression.

\noindent\textit{\underline{Challenge 3}: How can we design a minimal
video uplink pipeline that supports VLM-based egocentric HOI logging?}
This challenge arises because SoTA egocentric HOI understanding increasingly relies on
VLMs~\cite{xu2024egocentric}.
However, existing VLM-based pipelines typically assume access
to continuous RGB video.
In contrast, \sys produces \emph{disconnected} clips that are both
spatially cropped (RoI-centric) and temporally sparse, to reduce
on-device sensing and communication costs.
The key question is how to preserve HOI logging accuracy under
this constrained uplink \emph{without} fine-tuning the VLM,
while minimizing bandwidth.
To answer this, we systematically benchmark VLM-based HOI logging
across frame rate, resolution, compression level, and temporal
continuity, quantifying how each factor affects recognition accuracy.
Guided by these results, we design an HOI-aware video encoder that selects the
minimal set of RoI content, at the lowest feasible
resolution and compression rate, that still retains the essential
visual cues needed by off-the-shelf VLMs, to achieve performance
comparable to that of streaming continuous high-frame-rate,
high-resolution video.

We build a smartglasses prototype with a low-power event or monochrome camera, an RGB camera, and an IMU, and implement the \sys pipeline on a Raspberry Pi Zero with commodity BLE uplink.
We first evaluate \sys on the 50-hour Nymeria egocentric RGB benchmark~\cite{ma2024nymeria} with 97 subjects across diverse daily-life scenarios, simulating event frames from RGB frames. We assess HOI recognition and localization accuracy, average RGB frame rate, and uplink bandwidth requirements.
We then collect a 20-hour real-world dataset from 10 participants under IRB approval using our \sys prototype, comprising synchronized RGB, event camera, and IMU recordings.
On this real-world testbed, we evaluate the same metrics and additionally measure system energy consumption and end-to-end latency.
Compared with continuous RGB recording and on-device YOLO-based hand detection at 1~FPS, \sys reduces always-on backend power consumption by $5.8\times$ and $10.4\times$, respectively, with minimal HOI recognition accuracy loss across downstream tasks and VLM models.
\sys reduces uplink bandwidth by 2.9$\times$ to below 130~kbps, making stable BLE video streaming feasible for egocentric HOI logging.
\sys incurs negligible accuracy loss compared to uniform sampling across daily activities from HOI-intensive cooking to HOI-sparse leisure, and across motions from subtle HOIs such as pouring water to large HOIs such as sweeping the floor.
We summarize our contributions as follows:

\noindent $\bullet$ We present \sys, the first event-driven smartglass system for resource-efficient egocentric HOI logging, using an always-on event or low-cost monochrome camera and IMU.

\noindent $\bullet$ We design a lightweight on-device pipeline to use event frames for egomotion cancellation, HOI detection, and event-guided RGB triggering and uplink compression.

\noindent $\bullet$ We implement \sys on resource-constrained embedded hardware with BLE uplink and evaluate it through prototyping, multi-user multi-scenario real-world testbed, and large-scale dataset benchmarks.
Demos and examples are at \url{https://eventhoi.github.io}.

\noindent $\bullet$ Comprehensive experiments show that \sys reduces power and minimizes uplink bandwidth while maintaining HOI understanding accuracy.
\presec \section{Background and Related Work}
\label{sec:background-and-related-works}

\textbf{Egocentric HOI Understanding.} Egocentric HOI understanding uses head-mounted cameras to recognize how users manipulate and interact with objects.
It is important for modeling user behavior and enabling assistive systems to understand daily activities and intent.
Despite recent progress, this task remains challenging.
First, it requires reasoning about fine-grained hand-object relationships, subtle interaction cues, and \textit{diverse open-set} objects, which cannot be captured by conventional closed-set detection alone~\cite{shan2020understanding,wu2024exploring}.
Frequent head motion and the limited field of view further complicate robust perception in egocentric videos~\cite{xu2024egocentric,pei2025egothinker}.
Second, SoTA HOI understanding often relies on computationally intensive models, such as ResNet-based backbones, transformers, and VLMs~\cite{liu2022joint,chen2023detecting,radevski2023multimodal,zhang2023helping,bao2024handsonvlm,perrett2025hd}.
These models are difficult to deploy on resource-constrained wearables and often require transmitting data to the cloud, incurring substantial computation and communication overhead.
In contrast, \sys addresses egocentric HOI understanding from a systems perspective by reducing sensing, computation, and communication costs on wearable devices.

\begin{figure}[t]
    \centering
    \begin{tikzpicture}
        \node[anchor=south west, inner sep=0] (img) {\includegraphics[width=0.65\columnwidth]{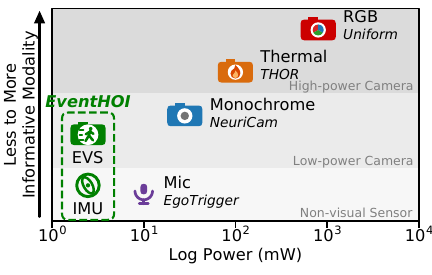}};
        \begin{scope}[x={(img.south east)}, y={(img.north west)}, every node/.style={anchor=base west, inner sep=0, font=\fontsize{5.5}{6}\selectfont}]
            \node at (0.6998, 0.7102) {\cite{shahi2025thor}};
            \node at (0.6501, 0.5356) {\cite{veluri2023neuricam}};
            \node at (0.5566, 0.2511) {\cite{paruchuri2025egotrigger}};
        \end{scope}
    \end{tikzpicture}
    \vspace{-0.2in}
    \caption{Comparisons of sensors in baseline methods used to reduce RGB camera usage.}
    \vspace{-0.2in}
    \label{fig:sensor-energy-comparison}
\end{figure}

\newcommand{\cmp}[1]{{\scriptsize\textcolor{red}{(#1)}}}
\begin{table}[t]
  \centering
  \caption{Power consumption and expected battery life. Red indicates comparisons to the first-row idle state.}
  \vspace{-0.1in}
  \label{tab:workload-power}
  \footnotesize
  \setlength{\tabcolsep}{0.5pt}
  \resizebox{\columnwidth}{!}{%
  \begin{tabular}{ccccc}
    \toprule
    \multicolumn{2}{c}{\textbf{Workload}} &
      \multirow{2}{*}{\shortstack{\textbf{Power}\\\textbf{(W)}}} &
      \multirow{2}{*}{\shortstack{\textbf{Battery}\\\textbf{Life (hrs)}}} &
      \multirow{2}{*}{\textbf{Use Cases and Examples }} \\
    \cmidrule(lr){1-2}
    Radio & Sensor & & & \\
    \midrule
    Off  & Off & 0.1                    & 7.5              & No video capability, no stream, on-device only \\
    BLE  & Off & 0.15~\cmp{$1.5\times$} & 5.0~\cmp{$67\%$} & Low-resource modality stream (image, audio, etc.) \\
    Off  & RGB & 0.7~\cmp{$7\times$}    & 1.1~\cmp{$15\%$} & No real-time video stream\\
    WiFi & RGB & 1.1~\cmp{$11\times$}   & 0.7~\cmp{$9\%$}  & Video stream (Ray-Ban Meta~\cite{RayBanMetaGen2}, Project Aria~\cite{engel2023project})\\
    \bottomrule
  \end{tabular}%
  }
  \vspace{-0.1in}
\end{table}

\noindent
\textbf{Event-based Computer Vision.} Event cameras, also known as event vision sensors (EVS)~\cite{mueggler2015lifetime,brandli2014real}, are low-power bio-inspired sensors that asynchronously record changes in pixel intensity rather than capturing full image frames at fixed rates.
Because they naturally respond to motion, they have been widely used in motion-centric tasks such as image deblurring, video reconstruction, drone perception, and autonomous driving~\cite{chen2020event,jiang2020learning,scheerlinck2020fast,cao2024eventboost,kong2025eventfly}.
More recently, they have also been explored in wearable systems for gesture recognition, object detection, and OCR~\cite{bhattacharyya2024helios,moosmann2024ultra,brander2025reading}.
Most prior systems rely on event frames, which aggregate asynchronous events into frame-like representations.
However, existing work mainly focuses on spatial perception, using events to recover appearance or semantic cues.
Their temporal dynamics remain underexplored for continuous reasoning tasks such as egocentric HOI understanding.
\sys instead exploits the motion sensitivity of event frames to trigger the RGB camera selectively and enable lightweight HOI detection and HOI-aware video compression.

\noindent
\textbf{Semantic Video Compression.}
Semantic video compression reduces bandwidth by prioritizing task-relevant content, such as salient objects, important regions, or meaningful events~\cite{hojjat2024limitnet}.
Existing approaches typically rely on additional saliency estimation modules, such as caches or neural networks, which introduce nontrivial computation, memory, and energy overhead~\cite{hojjat2024limitnet, hojjat2025mcucoder, xiao2024task,wang2023semantic}.
This overhead is particularly undesirable on wearable devices.
Recent multimodal compression systems further improve efficiency by combining heterogeneous sensors (Fig.~\ref{fig:sensor-energy-comparison}).
THOR~\cite{shahi2025thor} uses an always-on thermal camera to detect the presence of hands but still incurs high energy consumption ($\sim$150mW)~\cite{flir2019lepton25}.
EgoTrigger~\cite{paruchuri2025egotrigger} applies sound cues as indirect non-visual evidence to trigger RGB samples but is prone to failure in noisy environments. The close-set prediction model is also difficult to generalize to the diverse real-world acoustic environments.
NeuriCam~\cite{veluri2023neuricam} continuously captures a low-resolution grayscale stream and periodically acquires high-resolution color key frames, from which it reconstructs high-resolution color video at the receiver.
However, such designs still depend on periodic RGB capture and therefore cannot fundamentally reduce sensing cost.
In contrast, \sys uses event motion derived from an ultra-low-power IMU and an ultra-low-power event camera (<10mW for each) to trigger RGB capture selectively and guide lightweight HOI-aware encoding, reducing sensing and uplink overhead.
\presec
\section{Motivation and Preliminary Study}\label{sec:motivation}

\presec
\subsection{HOI Logging Pipeline and Constraints}
Most wearable egocentric HOI logging systems follow the same high-level pipeline.
First, a head-mounted RGB camera continuously captures egocentric video.
Second, the wearable device processes and compresses the frames.
Third, the compressed stream is transmitted to a smartphone or edge/cloud backend.
Finally, a backend CV-based model performs HOI understanding and converts the results into structured activity logs.
While backend inference can be offloaded, the front-end stages, i.e., RGB sensing and wireless transmission, must run continuously on the smartglasses and therefore dominate their energy and thermal budget.
We therefore focus on two coupled bottlenecks:

\noindent\textbf{(1) Camera sensing/recording energy.}
Modern smartglasses such as Ray-Ban Meta~\cite{RayBanMetaGen2} and Project Aria~\cite{engel2023project} support high-quality video capture, but continuous RGB sensing remains power-hungry.
Commodity RGB sensors typically consume 200--600\,mW depending on resolution, frame rate, and encoding settings~\cite{OG02B1B, OV5640}.
Given the limited battery capacity of smartglasses (e.g., $\sim$1--2.5\,Wh), continuous RGB recording only supports short sessions (tens of minutes
to $\sim$1.5 hours) rather than day-long operation~\cite{RayBanMetaGen2, engel2023project}.

\noindent\textbf{(2) Wireless uplink energy and bandwidth.}
Many smartglasses stream video at 720p and 15\,FPS~\cite{aria_gen1_specs, RayBanMetaGen2} by default, which typically requires 2--5\,Mbps depending on compression quality.
Such bitrates often require Wi-Fi rather than low-power radios such as BLE, adding substantial communication energy overhead.
It makes continuous video offloading impractical under the battery budget of smartglasses.
These tradeoffs motivate our preliminary studies.

\begin{figure}[t]
    \centering
    \includegraphics[width=0.9\columnwidth]{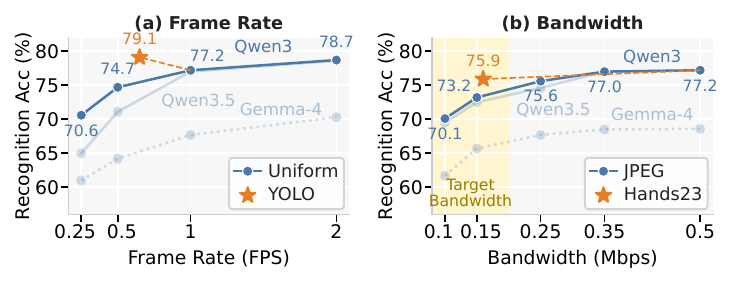}
    \vspace{-0.2in}
    \caption{Preliminary study on VLM-based HOI understanding for frame rate and uplink bandwidth.}
    \label{fig:preliminary}
    \vspace{-0.2in}
\end{figure}

\begin{figure*}[t]
    \centering
    \vspace{-0.1in}
    \includegraphics[width=0.8\textwidth]{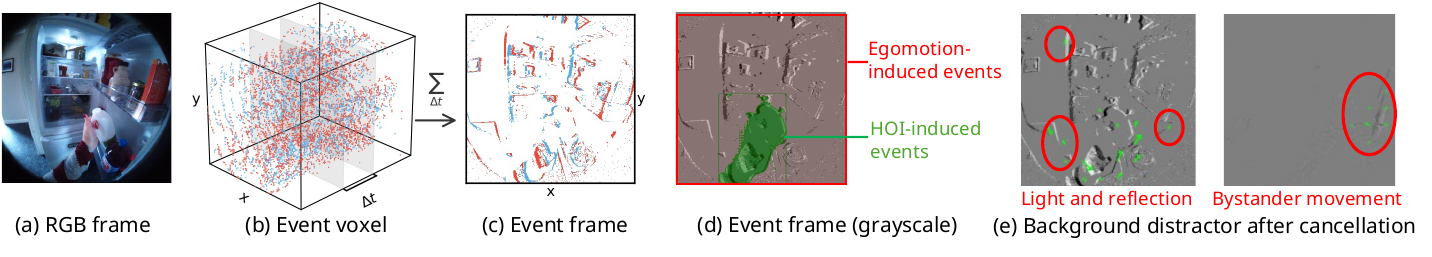}
    \vspace{-0.25in}
    \caption{(b) Asynchronous event voxels are aggregated to (c) an event frame (red for positive and blue for negative events). (d) Egomotion-induced and HOI-induced events. (e) Background lighting and bystander distractors after egomotion cancellation. }
    \label{fig:voxel}
    \vspace{-0.1in}
\end{figure*}

\presec
\subsection{Study 1: On-device Power Breakdown}
Existing smartglasses prototypes do not expose fine-grained power traces.
To understand which front-end components dominate the energy budget, we conduct an on-device power study using an ESP32-S3 with an OV5640 camera~\cite{seeedstudioxiao}, representative of community smartglasses prototypes such as OpenGlass~\cite{basedhardware2025omi}.
Specifically, we use an nRF-PPK2~\cite{nordicppk2} to measure power
under four configurations:
(i) idle with no workload,
(ii) BLE workload on an nRF52840 at 0.8~Mbps using 1M PHY for comparison~\cite{seeedstudioxiao},
(iii) RGB camera active at 720p and 1~FPS,
and (iv) Wi-Fi enabled to offload the video stream at 0.8~Mbps.
Tab.~\ref{tab:workload-power} reports the measured power breakdown
and the estimated battery life for a 200\,mAh LiPo battery.
\emph{Finding (1): Continuous RGB recording is the dominant on-device energy bottleneck.}
Enabling continuous RGB recording increases power by
$\sim$0.6\,W.
\emph{Finding (2): Wi-Fi uplink adds non-negligible overhead.}
Streaming video over Wi-Fi further adds about 0.4\,W, reducing estimated battery life to only $\sim$0.7\,h.

These results suggest that day-long HOI logging on smartglasses requires
\emph{(i) reducing RGB camera energy via selective capture}; and
\emph{(ii) lowering bandwidth enough to use low-power radios (e.g., BLE) instead of continuous Wi-Fi streaming.
}

\presec
\subsection{Study 2: Frame Rate and Uplink Bandwidth} \label{sec:prelim-study-2}
Study~1 shows that continuous RGB recording and Wi-Fi streaming
are prohibitively expensive for day-long operation.
A natural alternative is to trigger the RGB capture selectively and
transmit a much smaller visual stream over low-power links such as BLE.
However, HOI logging is task-driven: excessive reduction may remove the fine-grained hand-object cues and
critical temporal transitions required by the backend model.
We therefore study the minimum visual stream required to preserve HOI understanding performance.

Starting from a dense-video baseline (720p, 2\,FPS), we reduce the stream along three axes:
(i) frame rate reduction (from 2 FPS to 0.25 FPS),
(ii) compression strength, resulting in
different uplink bitrates, and (iii) content-aware selection.
Because our frame rate is very low, we compress each frame independently using JPEG rather than video codecs that rely
on inter-frame prediction.
For content selection, we consider an \emph{oracle} that uses CV-based hand detection to keep only frames likely containing HOIs, using YOLO~\cite{khanam2024yolov11} and Hands23~\cite{cheng2023towards} as described in \S\ref{sec:eval:benchmark}.
For each configuration, we evaluate HOI logging using
the off-the-shelf backend VLM Qwen3-VL-8B~\cite{yang2025qwen3}, which
provides strong HOI understanding performance and is used primarily for evaluations in this paper, Qwen3.5-9B~\cite{qwen2026qwen35}, and Gemma-4-12B-it~\cite{gemmateam2026gemma4technicalreport}.
We report HOI recognition accuracy (see \S\ref{sec:eval:benchmark}) on a 10-hour subset of the Nymeria dataset~\cite{ma2024nymeria}.

Fig.~\ref{fig:preliminary} shows two key observations.
\emph{(i) Content-aware selection is highly effective.}
Careful content-aware selection preserves similar HOI recognition accuracy while significantly reducing frame rate and bandwidth.
\emph{(ii) Moderate temporal/spatial reduction is tolerable.}
Performance remains largely stable at 1\,FPS, but drops sharply beyond that point on all tested models.
\textit{At 1\,FPS and 720p, the stream can be compressed to about 0.35\,Mbps with little accuracy loss, whereas stronger compression substantially degrades performance.}

In contrast, our empirical measurements of smartglasses-to-smartphone BLE uplink show that the available bandwidth can be as low as 0.3\,Mbps on iPhones. Concurrent transmission of other multimodal signals may further constrain the bandwidth available for RGB streaming: for example, empirical measurements on Ray-Ban Meta Gen 1 glasses indicate an AAC-encoded audio bitrate of 0.12--0.13 Mbps per channel. \textit{These observations motivate HOI-aware RGB compression methods that minimize uplink bandwidth while preserving comparable HOI recognition accuracy.}
\presec
\section{System Design}

Our key insights lead to a concrete systems question: \emph{how can 
smartglasses determine when an HOI is likely and where it occurs,
without paying the cost of continuous RGB sensing and Wi-Fi uplink?}
\sys answers this question with an always-on, low-power visual \textit{sentinel} that uses \emph{event frames} to \emph{gate} RGB capture and \emph{guide} compression.

\presec
\subsection{Low-Power Event Frame Capture}
\label{sec:low-power-event-frame}
As shown in Fig.~\ref{fig:voxel}, an event frame is a 
pixel-wise summary of recent visual changes, constructed by aggregating 
asynchronous intensity-change events over a short time window 
$\Delta t$~\cite{maqueda2018event}.
It highlights \emph{where} and \emph{how strongly} motion and 
scene dynamics occur while largely suppressing static 
background content.
This makes event frames well suited for HOI gating: 
static scenes produce sparse signals, whereas HOIs generate localized events around motion boundaries.

A straightforward way is to use 
low-power vision sensors to produce event frames, 
including event cameras and monochrome cameras.
An event camera aggregates asynchronous events over a short window to generate an \emph{event frame}~\cite{prophesee2025genx320}.
When event cameras are not available, a monochrome camera can be a practical fallback, since recent monochrome sensors support an ``event-like'' mode that subtracts consecutive frames with similarly low energy cost~\cite{ST_VD55G1}.

\noindent\textbf{Why is event-frame capture is more energy-efficient than RGB capture?}
Event cameras typically consume less than $10$~mW, over $60\times$ lower than OV5640 RGB cameras~\cite{prophesee2025genx320}.
This stems from their change-driven readout: the sensor reports only 
per-pixel brightness changes rather than full-frame images, so readout 
and downstream processing scale with scene dynamics rather than fixed 
frame rate and resolution.
Monochrome sensors still perform frame readout, but avoid the costly RGB ISP pipeline, such as demosaicing and color correction.
In ``event-like'' mode, they can operate at around $35$~mW, achieving roughly $17\times$ lower power than OV5640 RGB cameras~\cite{ST_VD55G1}.
Since \sys only needs low-resolution event frames for gating and localization, the always-on sensing cost remains low.

\begin{figure*}[t]
    \centering
    \includegraphics[width=0.85\textwidth]{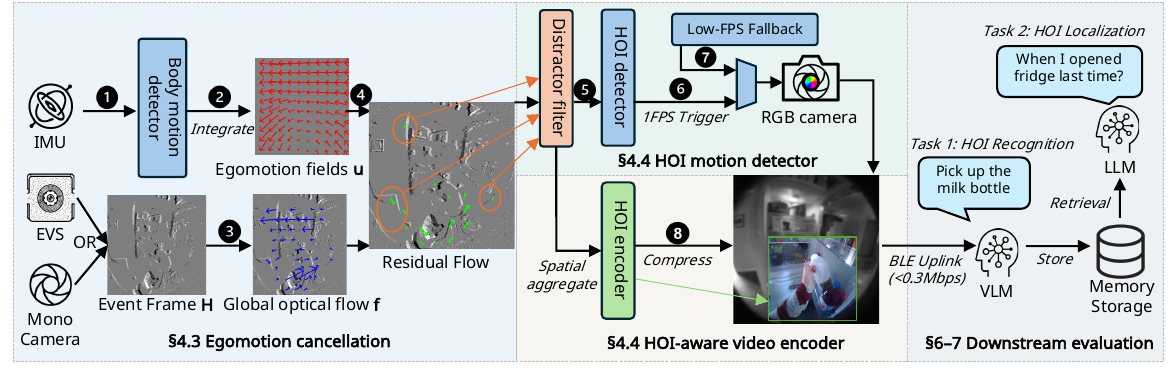}
    \vspace{-0.2in}
    \caption{EventHOI system overview.}
    \vspace{-0.1in}
    \label{fig:system-overview}
\end{figure*}

\presec
\subsection{Challenges and End-to-End Overview}
\label{sec:design-overview}
An event-frame sentinel is necessary, but not sufficient.
On head-mounted devices, non-HOI motion can dominate the event stream and cause frequent false triggers.
Fig.~\ref{fig:system-overview} summarizes how \sys addresses the following challenges.

\noindent\textbf{C1: Egomotion-induced background events.}
Head motion during walking or gaze shifts produces strong global changes across the field of view.
As shown in Fig.~\ref{fig:voxel}(d), these events can overwhelm HOI cues and keep the RGB camera unnecessarily active.
\sys uses the head-mounted IMU to estimate egomotion and cancel the corresponding global events in the event-frame stream (\S\ref{sec:system::egomotion}).

\noindent\textbf{C2: Non-HOI dynamic distractors.}
Motion from bystanders, background, and moving background objects (Fig.~\ref{fig:voxel}(e)) can also generate concentrated events, causing additional false triggers.
\sys applies a lightweight HOI motion detector to identify localized residual motion patterns and suppress false triggers from bystanders and background dynamics (\S\ref{sec:system::detector}).

\noindent\textbf{C3: Bandwidth-efficient uplink without losing HOI cues.}
Following selective RGB triggering, we aim to compress the captured frames to minimize BLE uplink traffic while preserving fine-grained HOI details required by downstream visual models. This enables sustained RGB streaming over BLE alongside data from other sensing modalities.
\sys uses an HOI-aware video encoder for saliency-guided spatial compression, reducing uplink bandwidth while preserving end-to-end HOI understanding performance (\S\ref{sec:system::detector}).

\presec
\subsection{Egomotion Cancellation}\label{sec:system::egomotion}

Egomotion denotes the camera motion relative to the surrounding 
environment~\cite{jayaraman2015learning}.
On head-mounted smartglasses, natural head and body movements can generate 
substantial background events, which interfere with HOI detection and may 
unnecessarily keep the RGB camera active.
As shown in Fig.~\ref{fig:voxel}(d), motions such as walking and head turns can produce dense non-HOI events that dominate the event frame.
Thus, the goal of this section is to design a lightweight, on-device
method to (1) estimate egomotion and (2) cancel the events 
induced by egomotion from the event frame.

\presec
\subsubsection{Egomotion Estimation.}

Accurate egomotion estimation is challenging on resource-constrained, 
battery-powered devices.
A head-mounted IMU available on most smartglasses offers a low-power
option (<$10$~mW) and has been widely used for human motion estimation~\cite{santaera2015low,forster2016manifold}.
In principle, linear displacement and angular rotation can be obtained 
by integrating linear acceleration and angular velocity, respectively.
However, integration error is the key challenge: displacement estimation requires double integration of acceleration, which amplifies sensor noise and bias and accumulates into large position drift over time~\cite{wang2015stance,norrdine2016step}.
Existing systems typically address this issue either by fusing camera and IMU 
signals, as in VIO and SLAM~\cite{forster2016manifold,campos2021orb}, or by 
using periodic zero-velocity references (ZUPT)~\cite{wang2015stance,norrdine2016step}.
However, camera fusion requires pixel-level matching on high-FPS video (e.g., 20--30~FPS)~\cite{campos2021orb,aria_gen1_specs}, adding sensing and compute overhead that conflicts with \sys's low-power goals.
On the other hand, ZUPT works well for foot-mounted IMUs with stance phases, but is unreliable for head-mounted IMUs that rarely become stationary~\cite{wang2015stance,norrdine2016step}.

Our key idea is to leverage prior knowledge about when informative HOIs 
are most likely to occur to guide egomotion estimation.
We observe that meaningful HOIs are rarely associated with large translational 
motion of the user, which often reflects whole-body movement rather than 
interaction-relevant behavior.
For example, activities such as walking can generate substantial global motion 
but seldom correspond to informative HOIs.
By contrast, HOIs typically occur when the body is relatively stable, as 
successful object interaction requires coordinated eye, head, and hand 
movements~\cite{li2013learning}.
Based on this observation, we decompose egomotion into two categories: 
\emph{(1) translational body motion}, such as walking, which is often accompanied
by few HOIs, and \emph{(2) rotational head motion}, such as small head turns
while standing or sitting, which commonly co-occurs with HOIs.

\noindent
\textbf{Step 1: Translational Body Motion Detector.} 
We first identify periods of large translational body motion, during which 
informative HOIs are unlikely~\cite{li2013learning}.
Based on our empirical study, we define such motion as linear velocity above 
$0.5$~m/s, since head motion alone rarely exceeds this threshold.
Rather than directly integrating IMU signals, we train a lightweight 2-layer 
LSTM binary detector that takes 6-axis IMU data sampled at $50$~Hz over a 
0.5-second window as input.
We train the model using the large-scale egocentric Nymeria dataset, collected 
with Meta Aria smartglasses~\cite{ma2024nymeria,aria_gen1_specs}.
Ground-truth labels are derived from SLAM-estimated egomotion velocities, with 
windows above $0.5$~m/s labeled positive.
To maintain class balance, we use a 5-hour indoor subset for training.
The detector runs with a default duty cycle of 0.5.
When large translation is detected, \sys directly disables RGB recording.
We adopt binary detection rather than precise velocity estimation because exact 
translation is unnecessary for HOI triggering and too costly for on-device processing.
Our model contains only 37k parameters (\textasciitilde37~KB of int8 weights) and
requires \textasciitilde1.8~MFLOPs per 0.5-second window, making it practical for low-power MCU deployment.
Despite this, our model shows a high prediction accuracy of 95.7\% and a recall of 0.96 for no-translation-period recognition, which minimizes the likelihood of skipping HOIs.

\begin{figure}[t]
    \centering
    \includegraphics[width=0.7\columnwidth]{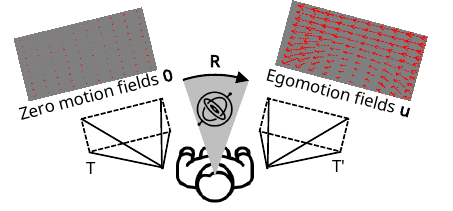}
    \vspace{-0.2in}
    \caption{Frame-to-frame egomotion fields $u$ (red arrows) derived
    from the rotation matrix $R$.}
    \label{fig:head-rotation}
    \vspace{-0.2in}
\end{figure}

\noindent
\textbf{Step 2: Rotational Head Motion Estimation using Gyroscope.}
After gating out large translational body motion, the remaining egomotion 
is primarily head-rotation-dominated~\cite{li2013learning}.
We estimate this component directly from gyroscope measurements.
Specifically, we integrate the 3-axis angular velocity between two consecutive 
event-frame timestamps $T$ and $T'$ to obtain a frame-to-frame rotation matrix 
$\mathbf{R}$, as illustrated in Fig.~\ref{fig:head-rotation}.
We then project $\mathbf{R}$ into a 2D 
egomotion field $\mathbf{u}$, which models the per-pixel displacement induced solely by rotational egomotion.
Formally, given two consecutive event frames $H_T$ and $H_{T'}$ and 
interleaving gyroscope samples $\{\boldsymbol{\omega}_{t_1}, 
\boldsymbol{\omega}_{t_2}, \cdots, \boldsymbol{\omega}_{t_n}\}$, where 
$T \le t_1 < \cdots < t_n < T'$, we compute 
the frame-to-frame motion field $\mathbf{u}$ as follows.

\noindent \emph{Step 2.1: Gyroscope Reprojection.}
Raw gyroscope signals sampled at $50$~Hz are 
first low-pass filtered with a $10$~Hz cutoff to suppress high-frequency IMU noise.
The filtered angular velocities are then re-projected into the camera 
coordinate frame using the extrinsic calibration matrix $\textbf{E}$ 
provided by the manufacturer as $\boldsymbol{\omega}^r_t = \textbf{E} \boldsymbol{\omega}_t$.

\noindent \emph{Step 2.2: Angular Velocity Integration.}
Each re-projected angular velocity $\boldsymbol{\omega}^r_t$ is integrated 
over the IMU sampling interval $\Delta t$ to obtain an incremental rotation 
vector $\boldsymbol{\theta}_t = \boldsymbol{\omega}^r_t \Delta t$. The incremental rotation is converted to a quaternion via the 
exponential map $q_t =
\Big[
\tfrac{\boldsymbol{\theta}_t}{\lVert \boldsymbol{\theta}_t \rVert}
\sin\!\big(\tfrac{\lVert \boldsymbol{\theta}_t \rVert}{2}\big),\;
\cos\!\big(\tfrac{\lVert \boldsymbol{\theta}_t \rVert}{2}\big)
\Big]
$.

\noindent \emph{Step 2.3: Rotation Integration.}
Then, we compose the incremental quaternions to obtain the frame-to-frame rotation:
\vspace{-0.1in}
\[
Q_{TT'} =
\prod_{T \le t_i < T'}^{\rightarrow} q_t
=
q_{t_1} \otimes q_{t_2} \otimes \cdots \otimes q_{t_n}.
\]
The resulting quaternion $Q_{TT'}$ is finally mapped to the 
frame-to-frame rotation matrix $\mathbf{R}_{TT'}$.

\noindent \emph{Step 2.4: Motion Field Generation.}
Given $\mathbf{R}_{TT'}$, we compute the corresponding 2D rotational 
motion field. 
For each pixel $\mathbf{p}=(x,y)^\top$ with homogeneous 
coordinate $\tilde{\mathbf{p}}=(x,y,1)^\top$, the rotation-induced 
pixel location $\mathbf{p}'$ is obtained via the homography
$\tilde{\mathbf{p}}' \sim \mathbf{H}\tilde{\mathbf{p}}$, $\mathbf{H} = \mathbf{K}\mathbf{R}_{TT'}\mathbf{K}^{-1}$,
where $\mathbf{K}$ denotes the camera intrinsic matrix. 
After dehomogenization from $\tilde{\mathbf{p}}'$ to $\mathbf{p}'$, 
the motion field at pixel $\mathbf{p}$ is $\mathbf{u}(\mathbf{p}) = \mathbf{p}' - \mathbf{p}$, which represents the pixel displacement induced solely by rotational 
egomotion, as shown by the red arrows in Fig.~\ref{fig:head-rotation}.

\presec
\subsubsection{Egomotion Cancellation.}
Next, we use the egomotion field $\mathbf{u}(\mathbf{p})$ 
to suppress rotation-induced events in the raw event frame.

\noindent
\textbf{Step 3: Optical Flow Estimation.} 
Event frames encode per-pixel intensity changes but not explicit pixel displacements over time. 
We therefore estimate global motion between consecutive event frames using optical flow. 
Since this runs on resource-constrained smartglasses, it must remain lightweight.
We adopt the classical lightweight Lucas--Kanade (LK) sparse optical flow 
method~\cite{bouguet2001pyramidal}.
For each tracked pixel $\mathbf{p}$, LK estimates the optical flow vector 
$\mathbf{f}(\mathbf{p})$ by minimizing the photometric error over a local 
window $\Omega(\mathbf{p})$:
\begin{equation*}
\vspace{-0.05in}
\mathbf{f}(\mathbf{p})
=
\arg\min_{\mathbf{v}}
\sum_{\mathbf{q}\in\Omega(\mathbf{p})}
\Big(
I(\mathbf{q}+\mathbf{v},\, T')
-
I(\mathbf{q},\, T)
\Big)^2 
\vspace{-0.03in}
\end{equation*}
where $I(\mathbf{q},T)$ denotes the event-frame intensity at pixel 
$\mathbf{q}$ and time $T$, and $T'$ is the timestamp of the next frame.
We choose the sparse LK algorithm for two reasons. First, event frames are highly sparse, with only 5--10\% of pixels active,
making sparse flow estimation more suitable than dense methods
\cite{horn1981determining,farneback2003two,sun2018pwc,teed2020raft}.
Second, we find that aggressively downsampled event frames still yield reliable global motion estimates, enabling sparse point tracking and substantially reducing computation.
We empirically choose to track fewer than $500$ points, which incurs minimal loss in motion accuracy.

\noindent
\textbf{Step 4: Egomotion Cancellation.} 
Finally, for each pixel $\mathbf{p}$, we combine the gyroscope-derived 
egomotion direction $\mathbf{u}(\mathbf{p})$ with the measured optical
flow vector $\mathbf{f}(\mathbf{p})$. 
Since $\mathbf{u}(\mathbf{p})$ encodes only the direction of rotation-induced
displacement, we suppress egomotion by removing from $\mathbf{f}(\mathbf{p})$ 
its component along $\mathbf{u}(\mathbf{p})$:
$\mathbf{f}_{\mathrm{res}}(\mathbf{p}) = \mathbf{f}(\mathbf{p}) - 
(\mathbf{f}(\mathbf{p})^\top \mathbf{u}(\mathbf{p})) \mathbf{u}(\mathbf{p})$.
The resulting residual flow $\mathbf{f}_{\mathrm{res}}(\mathbf{p})$ 
suppresses motion consistent with egomotion while preserving unexplained 
motion, which in egocentric settings is largely associated with HOIs.
The residual optical flow $\mathbf{f}_{\mathrm{res}}(\mathbf{p})$ is used for the next step.

\presec
\subsection{HOI Motion Detector \& HOI-Aware Encoder}\label{sec:system::detector}
After egomotion cancellation, the residual optical flow contains not only 
HOI-related motion but also non-HOI dynamic distractors such as 
background dynamics and bystander motion.
We aim to denoise residual flows into HOI saliency maps and use them 
as a lightweight HOI detector to trigger sampling of RGB frames.

\begin{figure}[t]
    \centering
    \includegraphics[width=0.8\columnwidth]{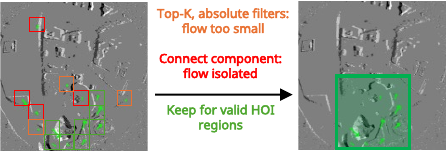}
    \vspace{-0.1in}
    \caption{Example of distractor filters to residual optical flow. Boxes in the left figure indicate the verdicts of filters. }
    \label{fig:filters}
    \vspace{-0.2in}
\end{figure}

\noindent
\textbf{Step 5: Spatial HOI Saliency Map Generation.}
We derive the HOI saliency map from $\mathbf{f}_{\mathrm{res}}(\mathbf{p})$ based on two observations from egocentric videos.
First, hand-induced motion usually has larger flow magnitude than background noise because hands are closer to the head-mounted camera.
Second, HOIs typically occupy a non-trivial image region due to the physical size of the hands and the typical hand-camera distance (e.g., 20--80\,cm).
Guided by these observations, we apply two denoising steps to suppress 
clutter and spurious motion.
First, we remove small residual flows with a magnitude-based filter: 
a pixel is retained only if its residual-flow magnitude exceeds an absolute 
threshold $\alpha$ and falls within the frame-adaptive top-$k$ percentile.
Second, we apply a spatial connectivity filter to remove tiny isolated clusters.
Specifically, we compute connected components on the binary mask and 
suppress components whose area is smaller than $C$\% of the image.
We refer to $C$ as the \emph{HOI-to-noise threshold}.
The resulting binary HOI saliency map $s$ at pixel $\mathbf{p}$ is
{
\[
s(\mathbf{p}) =
\begin{cases}
1, & \text{if } \|\mathbf{f}_{\mathrm{res}}(\mathbf{p})\|_2 \ge \alpha,\ 
\|\mathbf{f}_{\mathrm{res}}(\mathbf{p})\|_2 \in \text{top-}k, \\
& \text{and } \mathbf{p} \text{ belongs to a connected component}\ge C, \\
0, & \text{otherwise.}
\end{cases}
\]
}
Unless otherwise specified, we set $\alpha = 0.3$ and $k = 80\%$, which 
empirically provide a reasonable separation between meaningful HOI 
regions and background noise.
We illustrate an example of this process in Fig.~\ref{fig:filters}. 
We set $C=5\%$ by default, as determined by the ablation study in \S\ref{sec:eval:ablation-threshold}.

\noindent
\textbf{Step 6: Temporal HOI Detection.} 
We further exploit the temporal pattern of the HOI saliency map to avoid 
triggering RGB capture for visually uninformative periods.
In particular, stationary and low-motion HOIs, such as 
holding, resting, and pouring, often exhibit little motion and therefore contribute limited new visual information.
Thus, we design an HOI detector to only detect non-stationary HOI occurrences 
to achieve more efficient RGB frame sampling for HOI understanding.
Based on the denoised HOI saliency map $s$, we adopt a hysteresis-based sampling strategy to detect non-stationary hand motions as meaningful HOIs.
Specifically, we monitor whether $s(\mathbf{p}) = 1$ for any pixel $\mathbf{p}$ within the past $T=5$ seconds to enable RGB capture.
Our strategy enables efficient temporal-domain frame selection by 
mitigating the impact of momentary noise or brief motion dropouts. 
This prevents frequent on–off toggling of the RGB camera and, consequently, 
reduces temporal discontinuities that could confuse HOI understanding of backend VLMs.

\noindent
\textbf{Step 7: Adaptive Low-FPS Fallback Sampling.} Step 6 designs HOI detector for non-stationary HOIs. To capture stationary and low-motion HOIs that may not trigger the motion-based detector, e.g., a user holding a remote control for a long time, we add a 0.1~FPS fallback sampling mechanism if the detector is not triggered within a 10-second period. These frames will not be compressed by the encoder described in the following section. We evaluate the impact of \sys on this in \S\ref{sec:eval-stationary}.

\noindent
\textbf{Step 8: HOI-aware Video Encoder.} We first aggregate the HOI saliency maps $s$ over the past $T$ seconds and define the spatial cropping ROI as the smallest bounding box enclosing their union.
To avoid losing critical scene context, we enforce a minimum square RoI size of $S\%$ of the original frame.
If the resulting RoI is smaller than this threshold, we center it at the 
RoI centroid and symmetrically expand it until it reaches size $S\%$ on both dimensions.
This design is motivated by two properties of HOI video understanding.
First, existing video understanding models typically operate on uniformly 
sampled low-FPS frames and are sensitive to missing spatial context.
Cropping too tightly around the hands may exclude the manipulated object 
or surrounding scene cues needed to distinguish fine-grained interactions.
Second, because our motion-driven saliency map $s$ captures both hand and object motion, the resulting RoI naturally covers the most interaction-relevant 
regions.
This improves robustness to the temporal uncertainty of 
sparse RGB sampling, where captured frames may miss the exact contact moment.
By preserving both hand and object context, the encoder helps the downstream 
model infer the correct HOI even when RGB frames are sampled coarsely.
We use the same $T=5$~s as Step 6 since it can provide more spatial context by including the trajectory of hand movements for HOIs with a source or destination (e.g., from, to), like ``pick up milk bottle \emph{from the fridge}'', which cover over 12\% of our benchmark in \S\ref{sec:eval:benchmark}. We set the minimum retained frame size to $S=75\%$ by default based on the ablation study in \S\ref{sec:eval:ablation-framesize}.
These threshold-based solutions are extremely lightweight (32k FLOPs per second of input), allowing us to take advantage of egocentric perspective to select important information without using computationally heavy machine learning models like nano-size YOLO at 1 FPS (1.4G FLOPs per second of input)~\cite{khanam2024yolov11}. 
\begin{figure}[t]
    \centering
    \setlength{\abovecaptionskip}{-0.1mm}\includegraphics[width=0.9\columnwidth]{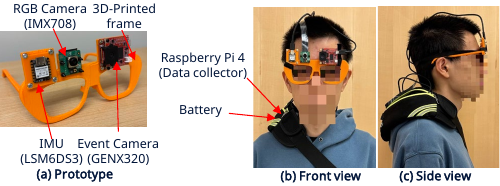}
    \caption{\sys prototype for data collection.}
    \label{fig:prototype}
    \vspace{-0.26in}
\end{figure}

\presec
\section{System Implementation}
\label{sec:implementation}
\noindent
We implement the \sys prototype on a 3D-printed smartglasses frame as shown in Fig.~\ref{fig:prototype}. We employ two different implementations for (1) continuous RGB, event, and IMU data collection (\S\ref{sec:testbed}) since continuous RGB capture is necessary for fair baseline comparisons and accuracy evaluations, and (2) minimized on-device system consumption measurements for reflecting real-world \sys efficiency on \textit{selective} RGB capture and compression (\S\ref{sec:system}).

\noindent
\textbf{(1) Data Collection Prototype.}
This setup is used to record complete RGB streams and event frames for the real-world testbed in \S\ref{sec:testbed}, so we prioritize comprehensive data collection over minimizing on-device resource consumption. We use an IMX708 RGB camera~\cite{arducam_b0474_imx708_datasheet_2023} at 720p and 30~FPS. We use a GENX320 event camera~\cite{prophesee2025genx320} in histogram mode to sample 10-FPS 320$\times$320 event frames. The two sensors are co-located for better alignment. The positions of the RGB and event cameras are further calibrated using a checkerboard target (with a toggling checkerboard for the event camera) and stereo rectification in OpenCV~\cite{opencv_library}. The cameras are connected to a Raspberry Pi 4B~\cite{raspberrypi4_datasheet_2024} for data collection. We use an nRF52840 MCU equipped with an LSM6DS3TR-C 6-DoF IMU~\cite{lsm6ds3tr} for head-mounted
IMU data collection~\cite{seeedstudioxiao}.
We use elastic headbands to securely mount the smartglasses prototype on the wearer's head. The smartglasses prototype weighs 46 grams, comparable to the weight of Ray-Ban Meta glasses~\cite{RayBanMetaGen2}. A sling bag carries the portable battery and Pi to minimize interference from the wearer's hand and body movements. Flex cables are used to connect the sensors on the smartglasses to the Raspberry Pi for data collection.

\noindent
\textbf{(2) On-device Implementation.}
This setup is optimized for measuring the actual on-device system consumption of \sys in \S\ref{sec:system}.
We implement all baselines on a resource-constrained Raspberry Pi Zero 2W SoC with a quad-core Cortex-A53 processor running at 600~MHz
and 512~MB of SDRAM~\cite{raspberrypi_zero2w_product_brief}, following~\cite{shahi2025thor}.
An ESP32-S3 microcontroller paired with an OV5640 RGB camera~\cite{seeedstudioxiao} operates
continuously in a low-power standby mode and wakes up only when triggered
by the HOI detector running on the Raspberry Pi Zero.
Other settings match (1) data collection prototype.
\presec
\section{Large-Scale Dataset Benchmark Evaluation} \label{sec:dataset-evaluation}
We evaluate whether the RGB uplink produced by \sys preserves
downstream HOI understanding quality while substantially reducing
the number of recorded frames and the stream transmission cost using a large-scale smartglasses dataset Nymeria~\cite{ma2024nymeria}.

\presec
\subsection{Benchmark Evaluation Setup} \label{sec:eval:benchmark}

\presec
\subsubsection{Evaluation Setups.}

\textbf{Dataset.}
We build our benchmark on the Nymeria dataset~\cite{ma2024nymeria}, which
contains egocentric RGB videos and head-mounted IMU recordings captured
with Project Aria glasses~\cite{engel2023project} across diverse scripted
daily activities.
Because Aria does not include an event camera, we simulate event frames at 10~FPS using the event simulator v2e~\cite{hu2021v2e} on RGB videos.
We study the simulation-to-real gaps to show the effectiveness of simulations in \S\ref{sec:eval:sim2real}.
The corresponding IMU signals are downsampled to 50~Hz.
These synchronized RGB, simulated event, and IMU streams are then used as
inputs to \sys and all baselines.

\begin{figure}[t]
    \centering
    \setlength{\abovecaptionskip}{-0.1mm}
    \includegraphics[width=\columnwidth]{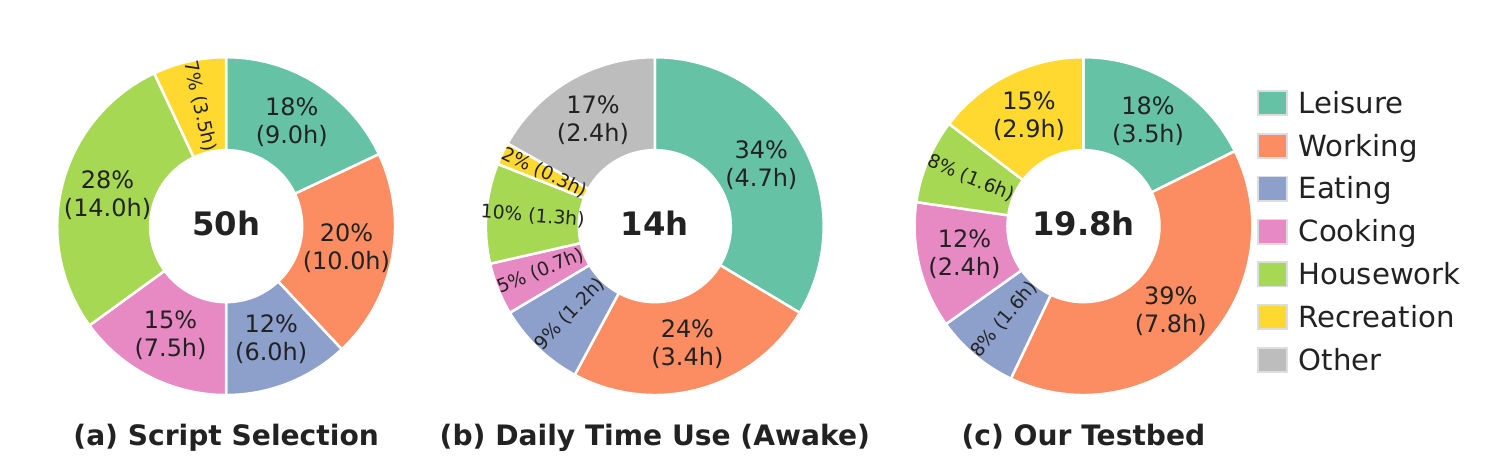}
    \caption{Nymeria~\cite{ma2024nymeria} sequence selection for evaluation, daily time-use survey results~\cite{BLS2025ATUS} (excluding sleeping time), and the recording time of our real-world testbed (\S\ref{sec:testbed}).}
    \label{fig:time-percentage}
    \vspace{-0.3in}
\end{figure}

\begin{figure}[t]
    \centering
    \includegraphics[width=0.8\columnwidth]{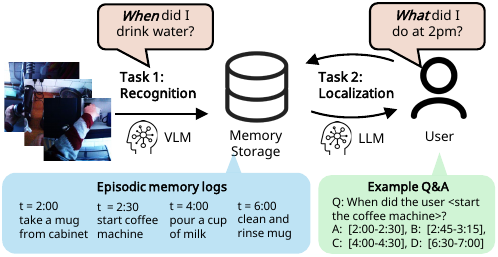}
    \vspace{-0.18in}
    \caption{Examples of HOI recognition and localization tasks for memory logging
    and retrieval.}
    \label{fig:evaluation-task}
    \vspace{-0.25in}
\end{figure}

To better reflect continuous daily-life logging, we construct a
representative 50-hour subset from Nymeria with 151 recording sequences and 97 subjects in 26 locations.
Specifically, we group sequences into 6 scenario categories (leisure,
working, eating, cooking, housework, and recreation) following the 14-hour
daily time-use distribution reported by the U.S.~Bureau of Labor
Statistics (Fig.~\ref{fig:time-percentage}(b))~\cite{BLS2025ATUS}.
We then weight the sequences of each category by its percentage share of the 14-hour daily time-use distribution.
Fig.~\ref{fig:time-percentage}(a) summarizes the resulting benchmark
composition.
We consider two downstream HOI understanding tasks, shown in
Fig.~\ref{fig:evaluation-task}: \emph{HOI recognition} and
\emph{HOI localization}.
These two tasks correspond to two core functionalities of egocentric
memory systems: generating text-based semantically meaningful HOI logs from RGB inputs and
retrieving relevant moments from those logs.

\noindent
\textbf{Task 1: HOI Recognition}
evaluates whether an algorithm-optimized RGB uplink stream can support accurate
transcription of user HOIs into natural-language narrations.

\noindent \emph{Benchmark Construction.}
Nymeria provides 5-second atomic activity annotations, but these labels are broader than the HOI-focused narrations needed here.
We therefore use Gemini-2.5-Pro~\cite{comanici2025gemini} as an annotation
oracle to convert each atomic label into an HOI-oriented natural-language
narration, such as ``pick up a water bottle'' or ``hold a cup under the
dispenser.''
To avoid annotation drift or hallucination, we manually verify the oracle
outputs and use them as ground truth.
We curate more than 22,000 HOI annotations from the 50-hour benchmark.
For each annotation, we use the corresponding 5-second time window from Nymeria for evaluation.

\noindent \emph{Evaluation Protocol.}
For each method, we feed its resulting RGB uplink stream into a
state-of-the-art off-the-shelf VLM to generate HOI narrations.
We then compare each generated narration against the corresponding
ground-truth HOI annotation.

\noindent \textbf{Task 2: HOI Localization}
evaluates whether \textit{timestamped natural-language HOI logs} from Task 1 can support
episodic retrieval, i.e., whether an LLM can localize when a queried HOI occurred.

\noindent
\emph{Benchmark Construction.}
For each $\sim$20-minute video, we use GPT-5.2~\cite{singh2025openaigpt5card}
to group the 5-second ground-truth annotations from Task~1 into 15--30-second
activity summaries and generate 10--20 temporal retrieval queries.
Following standard video-QA protocols
\cite{mangalam2023egoschema, xiao2021next, fu2025video}, each query is paired
with one correct time window and three distractors from the same video,
yielding 1,533 multiple-choice questions.

\noindent
\emph{Evaluation Protocol.}
For each question, the answering LLM is given the full-video memory logs,
i.e., the VLM narrations generated from sampled frames in Task~1, and selects
the correct window from the four choices.
This evaluates whether the sampled RGB frames preserve enough semantic evidence
for accurate HOI localization.

\presec
\subsubsection{Evaluation Metrics.}\label{sec:metric}

\textbf{Metric 1: HOI recognition accuracy.}
Qwen3-VL-8B~\cite{yang2025qwen3} VLM is primarily used in the HOI recognition evaluations according to its strong performance shown in \S\ref{fig:preliminary}. We further supply results on the state-of-the-art Qwen3.6-27B~\cite{qwen2026qwen35} and Gemini-3-Flash~\cite{google2025gemini3promodelcard} VLMs to show the generalization of \sys.
For Task~1, we compute the sentence similarity between each generated
narration and its ground-truth annotation using the BGE-M3 sentence
embedding~\cite{chen2025m3embedding}.
Following prior semantic matching practice~\cite{reimers2019sentence}, we binarize the score with a threshold of $0.65$~\cite{zhang2018fine,bansal2022sem}.

\noindent \textbf{Metric 2: HOI localization accuracy.} For Task~2, we report accuracy over all 1,533 multiple-choice QAs,
using both GPT-5.2~\cite{singh2025openaigpt5card} and Gemini-2.5-Pro~\cite{comanici2025gemini} with medium-level thinking as the answering LLMs to ensure
robustness across retrieval models.

\noindent \textbf{Metric 3: Effective Frame Rate.}
To quantify temporal RGB recording reduction, we define
\emph{effective frame rate} as the average number of RGB frames
sampled per second after temporal gating.
Unless otherwise specified, the default frame rate at triggering is 1~FPS based on
our preliminary studies (\S\ref{sec:motivation}), and at a fallback of 0.1~FPS otherwise (\S\ref{sec:system::detector}).

\noindent \textbf{Metric 4: Uplink Bandwidth Requirement.}
To measure actual transmission cost, we report the average uplink
bandwidth in Mbps.
Unless otherwise specified, the default resolution is 720p
(\S\ref{sec:motivation}).

\begin{table}[t]
\centering
\vspace{+0.1in}
\caption{Comparison of baselines and \sys ablations.}
\vspace{-0.15in}
\label{tab:baselines}
\setlength{\tabcolsep}{1.5pt}
\resizebox{\columnwidth}{!}{%
\begin{tabular}{lcccccc}
\toprule
& \multicolumn{3}{c}{\textbf{Baselines}}
& \multicolumn{3}{c}{\textbf{\sys Ablations}} \\
\cmidrule(lr){2-4}
\cmidrule(lr){5-7}

& Uniform & YOLO~\cite{khanam2024yolov11} & Hands23~\cite{cheng2023towards} & Vanilla EVS & Detector only & \sys \\
\midrule
Reduce RGB frames?
& \textcolor{red}{No}
& \textcolor{green!60!black}{Yes$^{*}$}
& \textcolor{red}{No}
& \textcolor{red}{No}
& \textcolor{green!60!black}{Yes}
& \textcolor{green!60!black}{Yes} \\

Can compress frames?
& \textcolor{red}{No}
& \textcolor{red}{No}
& \textcolor{green!60!black}{Yes}
& \textcolor{red}{No}
& \textcolor{red}{No}
& \textcolor{green!60!black}{Yes} \\

On-device feasible?
& \textcolor{green!60!black}{Yes}
& \textcolor{green!60!black}{Yes}
& \textcolor{red}{No}
& \textcolor{green!60!black}{Yes}
& \textcolor{green!60!black}{Yes}
& \textcolor{green!60!black}{Yes} \\
\bottomrule
\end{tabular}
}

\vspace{1pt}
\parbox{\columnwidth}{\footnotesize $^{*}$ We consider YOLO as a CV-based \textit{oracle}. It does not reduce RGB frames in reality.}
\vspace{-0.35in}
\end{table}

\presec
\subsubsection{Baselines.} \label{sec:baseline}
We compare \sys against three RGB sampling and uplink-stream baselines and two ablations (Tab.~\ref{tab:baselines}):
\noindent \textbf{(1) Our solution (\sys)}.

\noindent \textbf{(2) Uniform Sampling (Uniform)}: Frames are sampled at a fixed rate and transmitted uniformly.
This is the default solution but incurs high RGB sampling and transmission cost.

\noindent \textbf{(3) YOLO Hand Detection Oracle (YOLO)~\cite{khanam2024yolov11}}:
This baseline assumes an oracle-like on-device hand detector running
continuously.
We fine-tune YOLOv8n on 2,000 manually annotated hand bounding boxes from Nymeria, then apply the same HOI motion detector as \sys
(\S\ref{sec:system::detector}) on the detected RoIs.
YOLO serves as an oracle for temporal hand localization, but requires uniformly sampled RGB frames to detect hands and therefore cannot avoid the RGB sampling cost in reality.
Moreover, the hand-only detector does not localize interacting objects and thus cannot directly support our HOI-preserving uplink compression.

\noindent \textbf{(4) Hands23 HOI Detection (Hands23)~\cite{cheng2023towards}}:
This stronger oracle baseline requires uniform RGB sampling and
predicts both hand masks and interacted-object regions. It therefore can compress frames but cannot reduce RGB sampling. Also, its large ResNet-101 backbone disallows resource-constrained on-device deployment.

\noindent \textbf{(5) \sys w/o Egomotion Cancellation (Vanilla EVS)}.
This ablation removes the egomotion cancellation module
(\S\ref{sec:system::egomotion}) while retaining the HOI motion
detector (\S\ref{sec:system::detector}) and HOI-aware video
encoder (\S\ref{sec:system::detector}). The raw, unprocessed event frames are used to trigger RGB camera capture and encode the RGB frames.

\noindent \textbf{(6) \sys Detector Only.}
This ablation disables the HOI-aware video encoder
(\S\ref{sec:system::detector}).
We do not include a variant without the HOI motion detector alone
(\S\ref{sec:system::detector}) since the encoder depends on
the detector outputs.

\presec
\subsection{Benchmark Evaluation Results}

\begin{figure}[t]
    \centering
    \includegraphics[width=0.9\columnwidth]{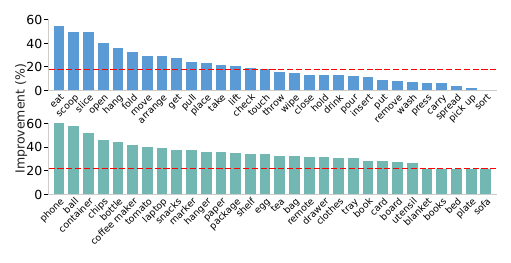}
    \vspace{-0.3in}
    \caption{Top-30 \sys (0.5~FPS effective frame rate) improvements over uniform sampling (0.5~FPS) on hand motions (blue) and object recognition (green). Red is the average. }
    \vspace{-0.2in}
    \label{fig:wordcloud}
\end{figure}

\presec
\subsubsection{Comparison to Uniform Sampling.}
We first compare \sys against uniform sampling to determine whether its gains come merely from recording fewer frames or from selecting more informative ones non-uniformly.
We use HOI recognition accuracy to quantify downstream task performance, effective frame rate to quantify sensing cost, and uplink bandwidth to quantify the transmission cost.
To better isolate the benefit of \sys, we exclude HOIs that are already
correctly recognized by uniform sampling at 0.25~FPS (discussed in \S\ref{sec:motivation}), as these are
predominantly long-duration interactions that remain observable even
when the RGB camera captures only one frame every four seconds.
This results in a subset of 6,644 challenging HOI queries.
\textbf{(1) Accuracy at the same effective frame rate.} For fairness, we set the HOI-to-noise threshold of \sys to $C=5\%$ (see ablation in  \S\ref{sec:eval:ablation-threshold}) to achieve the same 0.5~FPS \emph{effective frame rate}.
We observe a 9.0\% HOI recognition accuracy improvement on average across different scenarios compared to 0.5~FPS uniform sampling.
\textbf{(2) Accuracy at the same uplink bandwidth.}
We further observe a 12.2\% average accuracy improvement compared to 0.35~FPS uniform sampling at the same 0.12~Mbps BLE uplink bandwidth.
\textit{This result demonstrates the strong benefits of \sys's detector and encoder over uniform sampling at the same effective frame rate and BLE uplink bandwidth.}

\presec
\subsubsection{Analysis on Improvements to Uniform Sampling.}
To further understand the benefits of \sys compared to 0.5 FPS uniform sampling (the same effective FPS as \sys), we use the LLM-as-Judge method to analyze the HOI queries for which \sys and uniform sampling produce different predictions.
Using GPT-5.2~\cite{singh2025openaigpt5card}, we characterize whether each method correctly identifies (1) the object and (2) the hand motion.
We present the improvements for the top-30 objects and hand motions in Fig.~\ref{fig:wordcloud}, each of which appears in at least 1\% of the dataset. The results show that
(1) \sys improves object recognition by +22.1\% on average, with especially large gains (over +30\%) for small objects such as smartphones, containers, markers, and hangers, which are prone to occlusion and require multiple frames to understand;
and (2) \sys improves hand motion recognition by +17.8\% on average, with above-average gains (over +35\%) for short motions such as ``open (doors or cabinets)'', ``scoop or eat (food)'', and ``hang (clothes)'', which require high temporal resolution to track.
\textit{Overall, \sys is particularly effective for recognizing small objects and short-term HOIs. This observation is consistent with its ability to better temporally localize HOIs that are vulnerable to uniform downsampling. }

\presec
\subsubsection{Analysis on Stationary and Low-motion HOIs.} \label{sec:eval-stationary}
Beyond capturing fast motions, we evaluate whether \sys can retain evidences for stationary or low-motion actions, such as ``hold'', ``type'', ``write'', ``pour'', etc.
We use GPT-5.2 LLM to identify stationary HOIs by instructing the LLM to select queries of stationary or low-motion actions like above. This filtering yields a subset of 6,117 stationary HOI queries. At an effective frame rate of 0.5~FPS, \sys achieves similar HOI recognition accuracy (+1.9\%) compared to uniform sampling at the same frame rate, benefiting from the 0.1~FPS fallback sampling that captures near-stationary HOIs whose motion falls below the HOI detector threshold. \textit{This result demonstrates \sys's robustness in capturing subtle and stationary HOIs.}

\begin{figure}[t]
    \centering
    \includegraphics[width=0.95\columnwidth]{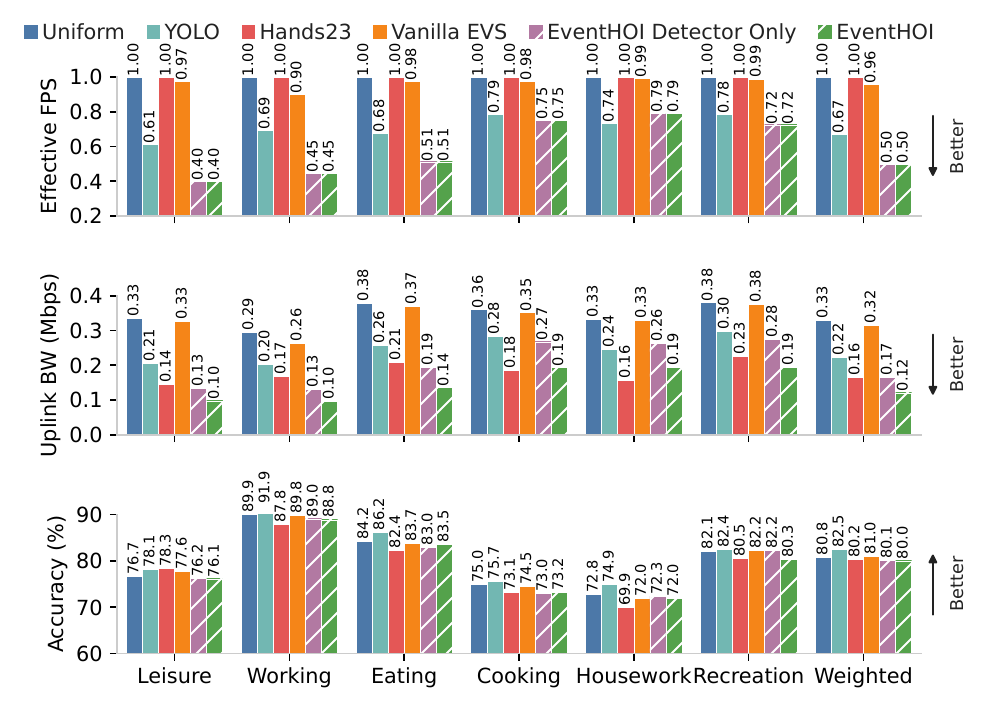}
    \vspace{-0.2in}
    \caption{Effective FPS, uplink bandwidth, and accuracy for baselines and \sys ablations on the Nymeria dataset~\cite{ma2024nymeria}.}
    \vspace{-0.15in}
    \label{fig:accuracy}
\end{figure}

\presec
\subsubsection{Comparison with CV-based Oracle Baselines.}
Next, we compare \sys against CV-based oracle baselines without considering on-device compute constraints.
The detected hand regions of YOLO is used for selective RGB sampling only since it cannot detect HOI-related objects, while
Hands23's HOI regions are used for HOI-aware compression strategy as \sys (\S\ref{sec:baseline}).
As shown in Fig.~\ref{fig:accuracy}, Hands23 requires
continuous RGB input and thus operates at the same effective frame rate
as uniform 1\,FPS sampling.
YOLO Oracle detects hand appearances at an effective frame rate of $0.67$\,FPS on average.
In contrast, \sys triggers RGB sensing only when HOI motion is detected,
reducing the effective frame rate to $0.5$\,FPS on average.
In uplink bandwidth, \sys substantially outperforms YOLO, since YOLO detects
only hands and cannot reliably support HOI-aware encoding.
\sys is also comparable to, and slightly better than, Hands23, whose
CV-based detections are not always robust in diverse egocentric scenes.
Meanwhile, \sys achieves HOI recognition accuracy comparable to both oracle baselines.
Overall, \emph{\sys matches the HOI recognition accuracy of CV-based oracle
baselines while requiring lower uplink bandwidth, without relying
on continuous RGB recording or heavyweight on-device CV inference}
(\S\ref{sec:system}).

\begin{figure}[t]
    \centering \includegraphics[width=0.7\columnwidth]{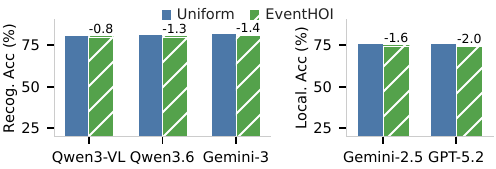}
    \vspace{-0.3cm}
    \caption{\sys recognition and localization 
    accuracy.}
    \label{fig:gemini}
    \vspace{-0.1in}
\end{figure}

\presec
\subsubsection{Performance across Different VLMs and Tasks.}
We further evaluate whether \sys generalizes across foundation models and downstream tasks.
For Task~1, \emph{HOI Recognition}, we compare the VLM-generated narrations produced by open-source Qwen3-VL-8B~\cite{yang2025qwen3}, Qwen3.6-27B~\cite{qwen2026qwen35}, and proprietary Gemini-3-Flash~\cite{google2025gemini3promodelcard} in different model sizes. 
Compared with the uniform sampling baseline, \sys incurs only negligible accuracy drops of 0.8\%, 1.3\%, and 1.4\%, respectively (Fig.~\ref{fig:gemini}).
For Task~2, \emph{HOI Localization}, we evaluate temporal HOI localization using two LLMs.
Compared with logs generated from the uniform sampling baseline, temporal localization accuracy using \sys-produced logs drops by merely 1.6\% for Gemini-2.5-Pro~\cite{comanici2025gemini} and 2.0\% for GPT-5.2~\cite{singh2025openaigpt5card}.
These results highlight two key findings.
First, \emph{\sys preserves the salient interaction evidence needed not only for HOI recognition, but also for downstream temporal reasoning in episodic memory retrieval.}
Second, \emph{the benefits of \sys's event-driven selective sampling are largely model-agnostic, yielding a consistent accuracy-efficiency tradeoff across different models.}

\presec
\subsubsection{Ablation on Components.}
We conduct two ablation studies, as described in \S\ref{sec:baseline}.
Without egomotion cancellation (Vanilla EVS, \S\ref{sec:system::egomotion}), \sys cannot reliably distinguish HOI-related motion from background egomotion, causing the effective frame rate to remain close to 1~FPS.
In contrast, the HOI motion detector (\S\ref{sec:system::detector}) and the HOI-aware video encoder (\S\ref{sec:system::detector}) further reduce the uplink bandwidth to below 0.2~Mbps across different activity scenarios with a negligible 0.1\% accuracy loss on average.
Overall, these results show that each design component of \sys is important to achieving the system's accuracy-efficiency tradeoff.

\presec
\subsubsection{Ablation on HOI-to-Noise Threshold.}\label{sec:eval:ablation-threshold}
\begin{figure}[t]
    \centering
    \includegraphics[width=0.9\columnwidth]{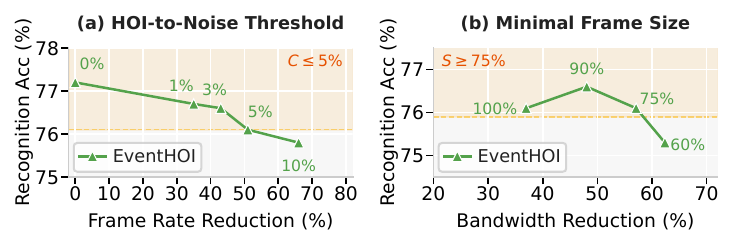}
    \vspace{-0.2in}
    \caption{Ablation studies on HOI-to-Noise threshold values for detector and minimal frame size for encoder. }
    \vspace{-0.3in}
    \label{fig:ablation-filter}
\end{figure}

We first determine the HOI-to-noise threshold $C$ of the HOI motion detector (\S\ref{sec:system::detector}),
since it critically balances preserving HOIs and
filtering distractors.
Specifically, we evaluate a 10-hour subset of the Nymeria
dataset~\cite{ma2024nymeria} using the HOI recognition task and the metrics
described in \S\ref{sec:eval:benchmark}, with a default sampling rate
of 1\,FPS.
We vary $C$ from 0\% to 10\%, and measure (i) the achieved frame-rate
reduction and (ii) the overall recognition accuracy in
Fig.~\ref{fig:ablation-filter}(a).
When $C=5\%$, \sys\ achieves about 50\% frame reduction while
maintaining recognition accuracy within 1\% compared to uniform 1\,FPS
sampling.
However, for $C>5\%$, accuracy drops sharply, indicating over-suppression of true HOI motion.
Thus, we set $C=5\%$ by default.

\presec
\subsubsection{Ablation on Minimum Frame Size.}\label{sec:eval:ablation-framesize}
We study the minimum retained frame size $S$ of the HOI-aware video encoder (\S\ref{sec:system::detector}) on 720p RGB frames on the same 10-hour subset.
Specifically, we fix the HOI-to-noise threshold at the default value
$C=5\%$ and vary $S$ from 60\% to 100\%, where larger $S$ preserves more
context but yields lower compression.
As shown in Fig.~\ref{fig:ablation-filter}(b), \sys maintains stable
recognition accuracy as $S$ decreases, until performance begins to drop
at $S=60\%$.
Based on this trade-off, we set $S=75\%$ by default, which keeps HOI
recognition accuracy within 0.3\% of using no HOI-aware encoder while
still achieving substantial spatial compression.

\presec
\subsubsection{Simulation-to-Real Gaps.}\label{sec:eval:sim2real}
Since the benchmark relies on v2e-simulated event frames, we validate whether the simulation-to-real gap affects the above results.
We evaluate over 2{,}000 synchronized RGB and real event frame segments from our real-world dataset benchmark (\S\ref{sec:testbed}). We compare optical flows computed from real event frames against those from v2e~\cite{hu2021v2e} events simulated on the paired RGB frames.
We then fit the measured angular error into a distribution, inject sampled errors into the simulated flows of the benchmark, and rerun the full \sys pipeline (setup details in Appx.~\ref{sec:appendix:sim2real}).
The results show that perturbed flows introduce negligible degradation to \sys: the selected frames maintain a high temporal overlap of 0.95 IoU with those selected from unperturbed flows, system consumption increases only slightly (+2.5\% effective frame rate and +5.1\% uplink bandwidth), and HOI recognition accuracy changes by only 0.2\%.
\textit{Overall, these results close the simulation-to-real gap and support the correctness of our large-scale simulation benchmark.}
\presec
\section{Real-World Testbed Evaluation}
\label{sec:testbed}
\presec
\subsection{Testbed Setup}

\presec
\subsubsection{Data Collection.}
We collect a 20-hour real-world dataset using the hardware described in \S\ref{sec:implementation} under IRB approval.
The dataset includes 10 participants (9 male and 1 female) in household, office, and open workspace (details in Appx.~\ref{sec:appendix:testbed}).
Participants are instructed to perform their daily activities naturally within one of the six scenarios with distributions in Fig.~\ref{fig:time-percentage}(c). We follow the weighted protocol of daily time-use in \S\ref{sec:eval:benchmark}.
We use synchronized RGB, event-camera, and IMU streams collected with the acquisition settings described in \S\ref{sec:implementation}.
Representative HOIs collected in our dataset are shown in Fig.~\ref{fig:testbed-wordcloud} in Appx.~\ref{sec:appendix:testbed}.

\presec
\subsubsection{Benchmark Setup.}
We use Gemini-3-Flash~\cite{google2025gemini3promodelcard} as a semi-automatic annotation assistant to generate HOI narrations for 5-second video windows.
Human annotators then verify and correct the generated narrations.
This process yields 1,950 HOI narrations for the evaluation of the HOI recognition task.
We follow the same evaluation protocol as described in \S\ref{sec:eval:benchmark}.

\presec
\subsection{Testbed Evaluation Results}

\presec
\subsubsection{Real-World Testbed Results.}
\begin{figure}[t]
    \centering
    \vspace{-0.1in}
    \includegraphics[width=1.0\columnwidth]{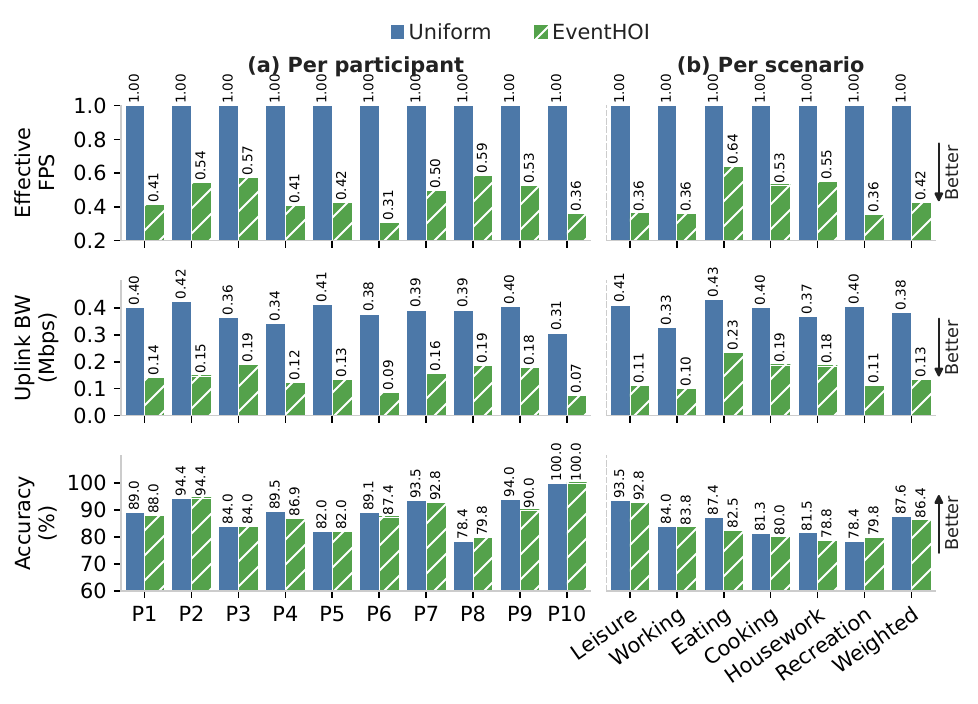}
    \vspace{-0.4in}
    \caption{Effective FPS, uplink bandwidth, and accuracy on the real-world testbed by participants (P1--10) and scenarios.}
    \vspace{-0.2in}
    \label{fig:accuracy-real-participant}\label{fig:accuracy-real}
\end{figure}

\begin{figure*}[t]
    \centering
    \includegraphics[width=0.86\textwidth]{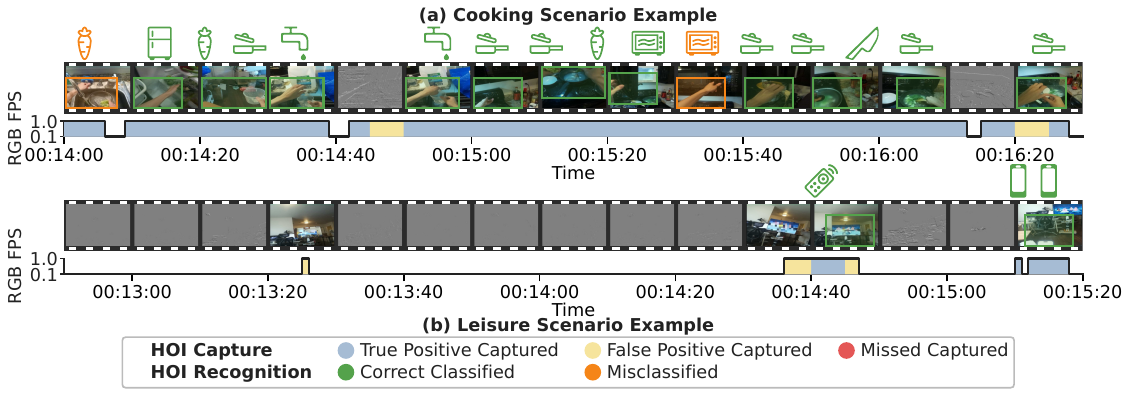}
    \vspace{-0.2in}
    \caption{\sys real-world examples. The plot shows the RGB sampling rate and the icons indicate different HOIs.}
    \vspace{-0.1in}
    \label{fig:real-benchmark-timeline}
\end{figure*}

We present per-subject and per-scenario testbed results for HOI recognition accuracy, effective frame rate, and uplink bandwidth in Fig.~\ref{fig:accuracy-real-participant}.
We observe \sys reduces the effective RGB frame rate by 2.4$\times$ (from 1 to 0.42~FPS) and BLE uplink bandwidth by 2.9$\times$ (from 0.38 to 0.13~Mbps) on average compared to the 1-FPS uniform sampling baseline. Despite these reductions, \sys incurs only a negligible 1.2\% degradation in weighted recognition accuracy.
Compared with the Nymeria benchmark~\cite{ma2024nymeria}, which focuses on densely recording HOIs, we observe the frame rate savings in real-world settings come from leisure, working, and recreation scenarios.
For common leisure activities such as watching TV, only 14.4\% of the windows contain HOIs (e.g., using a remote control). This allows \sys to skip 80.1\% of RGB frame sampling while still capturing these short HOIs when they occur.
Similarly, when participants use computers for working or recreation, we observe that over 59.1\% of the windows contain no HOIs in the camera view, as experienced users tend to look at the screen rather than their hands while typing. This allows \sys to reduce RGB frame sampling by 73.4\%.
We present examples from different scenarios in Fig.~\ref{fig:real-benchmark-timeline}.
\textit{Overall, our real-world testbed results demonstrate the substantially greater potential of \sys to reduce RGB sampling requirements in practical settings since our testbed follows more natural density of daily HOIs.}

\presec
\subsubsection{False Positives from Background Motion.}
In practice, we find that false-positive HOI detections primarily arise from background motion unrelated to either HOIs or egomotion, such as bystander movements and dynamic content displayed on TVs, smartphones, and monitors.
For example, dynamic TV content causes \sys to unnecessarily capture 3.8\% more windows in our testbed, while motion displayed on smartphones results in 1.2\% more captured windows.
Furthermore, movements from bystanders and pets result in an additional 0.3\% of captured windows.
This translates to less than 0.02~Mbps negligible uplink bandwidth increase and 20mW power increase on average (\S\ref{sec:system}). 
Although the HOI-to-Noise threshold effectively suppresses much of the background motion, these dynamic visual sources can still produce strong optical flow and be misclassified as HOI-related motion.
\textit{This limitation motivates incorporating more informative visual modalities, such as monochrome or RGB frames, to better distinguish true HOI saliency from background motion during HOI detection.}

\begin{figure}[t]
    \centering
    \includegraphics[width=0.8\columnwidth]{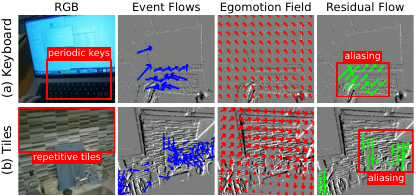}
    \vspace{-0.15in}
    \caption{Failure in flow aliasing on repetitive textures.}
    \label{fig:aliasing-failure}
    \vspace{-0.15in}
\end{figure}

\presec
\subsubsection{False Positives from Background Textures.}
We observe another failure mode caused by highly textured or repetitive backgrounds, including curtains, blinds, keyboards, bed sheets, and tiled surfaces, which lead to 1.7\% more captured windows in our testbed.
This is equivalent to less than 0.01~Mbps negligible uplink bandwidth increase and 10mW power increase on average (\S\ref{sec:system}). 
This behavior stems from a known limitation of LK-based optical flow: repetitive local patterns can create ambiguous correspondences, resulting in noisy or incorrect motion estimates~\cite{zhong2019unsupervised}.
As shown in Fig.~\ref{fig:aliasing-failure}, the periodic keys of a keyboard and the repetitive tiles of a backsplash produce aliased flow vectors pointing in opposite directions. Since egomotion induces a coherent motion field, egomotion cancellation cannot remove these opposite flows, and the residual is falsely detected as HOI saliency.
\textit{This limitation motivates future designs that combine event optical flow with lightweight appearance or semantic cues to better suppress false triggers from repetitive background textures.}

Despite these limitations, \sys consistently achieves substantial reductions in RGB frame sampling and uplink bandwidth across different real-world scenarios.
\textit{The real-world evaluation results demonstrate \sys's practicality in real-world settings.}
\begin{figure}[t]
    \centering
    \includegraphics[width=0.95\columnwidth]{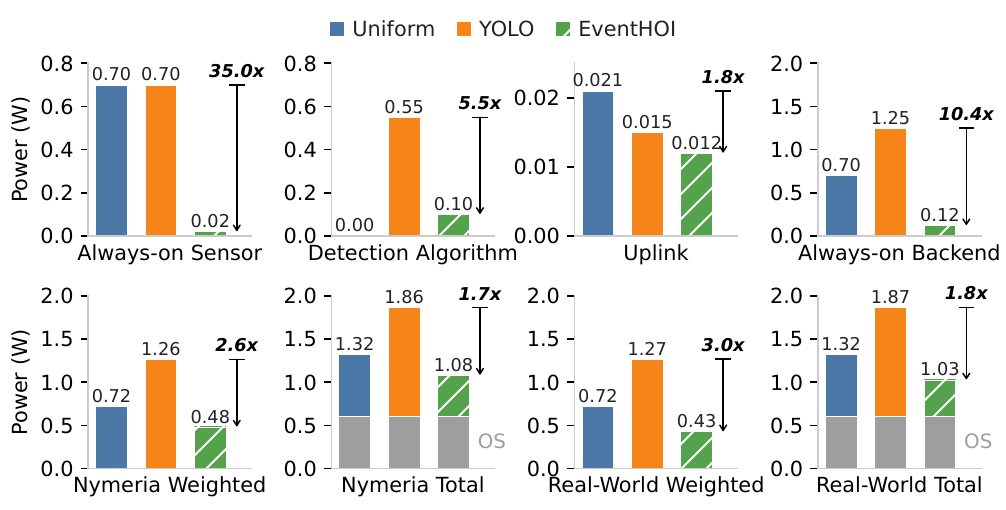}
    \vspace{-0.2in}
    \caption{Power consumption breakdown.}
    \label{fig:power-evaluation}
    \vspace{-0.2in}
\end{figure}

\presec
\section{System Evaluation}\label{sec:system}

We evaluate two key aspects of \sys’s on-device performance: (1) power consumption, which measures the computational efficiency of \sys, and (2) on-device latency, which evaluates whether the detection pipeline can operate in real time while meeting the RGB sampling duty cycle requirements. We assume the cloud backend VLMs are sufficiently powerful to process uploaded RGB frames in a timely manner.

\vspace{-0.15in}

\subsection{On-Device Power Consumption}
We compare the on-device power consumption of \sys against uniform sampling and YOLO. 
For a fair comparison, all methods are implemented on the on-device
prototype mentioned in \S\ref{sec:implementation}.
\sys~operates at 2~FPS for translational body-motion detection,
at 10~FPS for rotational head motion tracking, while triggering
the RGB camera at 1~FPS.
In contrast, all baselines run at 1~FPS, since they rely on
appearance-based detection and do not require finer-grained temporal sensing.

Fig.~\ref{fig:power-evaluation} shows the average power consumption
of each baseline.
\emph{\sys reduces the always-on sensing power by 35.0$\times$ compared
to Uniform and YOLO.}
\sys requires only $0.02$~W for continuous event-camera and IMU sensing,
whereas Uniform and YOLO both require continuous RGB capture, which
consumes $0.7$~W.
\emph{\sys also reduces the detection-stage power by 5.5$\times$
compared with YOLO.}
Although \sys introduces additional compute overhead for the HOI motion
detector (\S\ref{sec:system::detector}) and the HOI-aware video encoder
(\S\ref{sec:system::detector}), this overhead is only about $0.1$~W,
which remains small relative to the cost of continuous RGB sensing
($0.7$~W) and YOLO-based hand detection ($0.55$~W).
\emph{\sys reduces total uplink wireless communication power consumption
by 1.3--1.8$\times$ compared with Uniform and YOLO.}
\sys requires substantially lower uplink bandwidth and therefore
consumes only $0.012$~W for BLE.

\emph{Overall, during non-HOI periods, \sys reduces always-on backend
power by 5.8$\times$ and
10.4$\times$ compared with Uniform and YOLO, respectively.}
When mapped to the daily usage statistics in Fig.~\ref{fig:time-percentage}, this translates to \textit{33\% and 41\% energy savings on Nymeria and the real-world testbed, respectively, excluding the OS idle power consumption of Raspberry Pi OS.}

\presec
\subsection{On-Device Latency}
\noindent \textbf{HOI Detection Algorithm Latency.} Our results show that YOLO requires 6.93~s per
256$\times$256 frame and 1.92~s per 128$\times$128 frame on the RPi Zero, failing to localize hands reliably at lower resolutions.
Such high latency makes YOLO impractical even at 1~FPS.
In contrast, \sys processes each event frame in 15--20~ms, while the HOI
motion detector adds about 50~ms per 0.5~s window.
Even when event frames are processed at 10~FPS, \sys incurs only a
25--30\% duty cycle when driving 1-FPS RGB sampling.
Overall, \emph{\sys can decide 1-FPS RGB capturing within
100~ms.}

\noindent \textbf{End-to-end Latency.} \sys end-to-end on-device pipeline includes the HOI motion detector, RGB camera wake-up, RGB frame
capture, and camera's return to sleep.
We measure two metrics: (1) \emph{detection-to-capture latency}, from the
latest event/IMU input to the RGB trigger; and (2) \emph{RGB duty-cycle latency},
from waking the camera to putting it back to sleep.
The HOI detection algorithm takes less than 100~ms,
since event frames are processed in a streaming pipeline. Waking the ESP32-S3
from ultra-low-power sleep takes up to 300~ms. The default ISP pipeline on the
ESP32-S3 with the OV5640 camera performs automatic calibrations (e.g., white
balance and exposure), which can take up to 1~s to converge and is too slow
to meet our 1~FPS duty-cycle constraint.
To address this, we implement a checkpoint-and-restore mechanism that
stores the most recent camera calibration parameters in non-volatile
memory and restores them across power cycles, assuming the user remains
in similar lighting conditions across adjacent frames. This reduces
parameter convergence to 150--200~ms. Finally, checkpointing parameters
and shutting down the camera incur an additional $\sim$50~ms.

Overall, the worst-case detection-to-capture reaction latency is
about 500~ms, which
is acceptable because \sys aggregates saliency maps over a 5~s window and is
therefore tolerant to $\sim$0.5~s temporal staleness.
The RGB duty-cycle latency is 500--600~ms, satisfying the $<1$~s requirement.
\presec
\section{Conclusion}
We presented \sys, an event-driven system for resource-efficient egocentric HOI logging on smartglasses.
It uses low-power event sensing and IMU-assisted egomotion cancellation to selectively capture HOI-relevant RGB content without continuous RGB recording or heavyweight on-device computation.
More broadly, \sys points to a promising design paradigm for resource-constrained systems: using low-power sensors to guide and sparsify the operation of high-power sensors to enable resource-efficient embedded sensing systems.

\appendix
\section*{Appendix}

\section{Simulation-to-Real Evaluation}\label{sec:appendix:sim2real}

\begin{figure}[t]
    \centering
    \includegraphics[width=0.5\columnwidth]{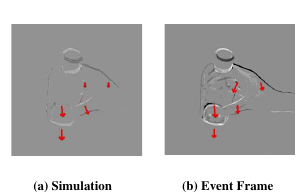}
    \caption{Optical flows (red arrows) from (a) v2e-simulated and (b) real event frames while the user holds a water bottle.}
    \label{fig:sim2real}
\end{figure}
This appendix details the setup of the simulation-to-real validation summarized in \S\ref{sec:eval:sim2real}.

\noindent
\textbf{Data Collection.}
We evaluate the 2{,}000 synchronized RGB and event frame segments from our real-world benchmark recording (\S\ref{sec:testbed}). The RGB and event camera hardware setup is discussed in \S\ref{sec:implementation} (1) Data Collection Prototype. 
We use the RGB frames to simulate events via v2e~\cite{hu2021v2e}, following the same simulation procedure as the benchmark in \S\ref{sec:eval:benchmark}.

\noindent
\textbf{Flow Error Measurement.}
For both real and simulated event frames, we compute optical flow at 10~FPS and 320$\times$320 resolution using the LK method (\S\ref{sec:system::egomotion}), and average the resulting flow over 8$\times$8 grids.
We remove outliers by discarding the largest anomalous flows (LK failures) and the smallest 10\% flows (noise).
We then compute the angular error between each pair of corresponding simulated and real grid flows, with examples shown in Fig.~\ref{fig:sim2real}.

\noindent
\textbf{Error Injection on the Benchmark.}
We fit the resulting angular errors into a circular normal distribution~\cite{fisher1995statistical}.
We then sample from this distribution and inject the errors into the optical flows produced by the RGB-v2e pipeline on each benchmark frame.
The perturbed flows are processed through the full \sys pipeline, including egomotion cancellation, HOI motion detection, and HOI-aware video encoding, and we generate HOI narrations using a VLM.

\noindent
\textbf{Evaluation Metrics.}
We compare the outputs of \sys on perturbed and unperturbed flows using the following metrics: (1) \textit{Temporal IoU} for the overlaps of selected frames, (2) \textit{Effective FPS}, (3) \textit{Uplink Bandwidth}, and (4) \textit{HOI Recognition Accuracy}.

\noindent
\textbf{Results.}
The simulation-to-real experiments reveal a median absolute angular error of 10--11 degrees with examples in Fig.~\ref{fig:sim2real}(a) and (b).
These errors likely arise from (1) stereo calibration and rectification inaccuracy, (2) v2e simulation mismatch, and (3) event sensor noise that cannot be fully removed by simple filtering.
Despite these error sources, the perturbed flows introduce negligible degradation to \sys: the selected frames maintain a high temporal overlap of 0.95 IoU, system consumption increases only slightly by +2.5\% effective frame rate and +5.1\% uplink bandwidth, and HOI recognition accuracy changes by only $-0.2\%$. 
\textit{Overall, these results close the simulation-to-real gap and support the correctness of our large-scale simulation benchmark.}

\section{Real-World Testbed Evaluation}\label{sec:appendix:testbed}
\begin{figure}[t]
    \centering
    \includegraphics[width=\columnwidth]{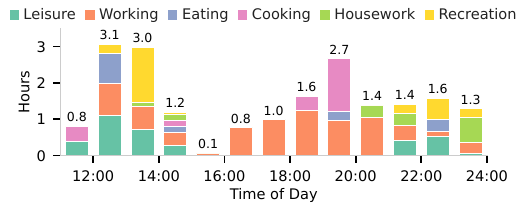}
    \vspace{-0.3in}
    \caption{Recording hours of the real-world testbed across the time of day.}
    \vspace{-0.1in}
    \label{fig:testbed-time-of-day}
\end{figure}

\begin{figure}[t]
    \centering
    \includegraphics[width=\columnwidth]{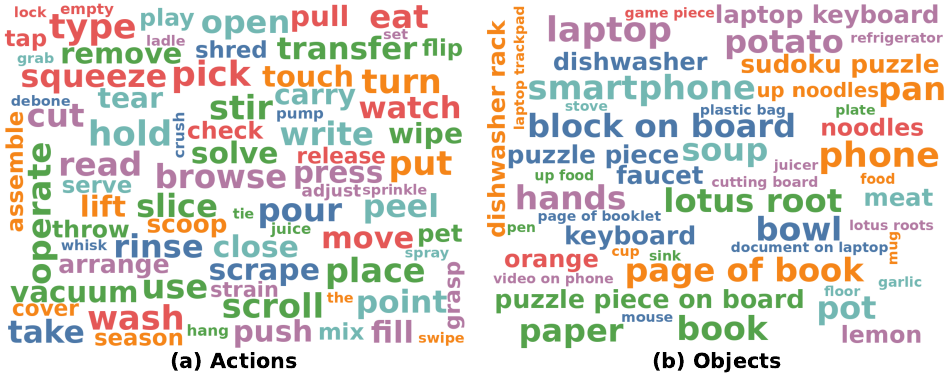}
    \vspace{-0.15in}
    \caption{Examples of HOIs in the real-world testbed.}
    \vspace{-0.15in}
    \label{fig:testbed-wordcloud}
\end{figure}

This appendix details the setup of the real-world testbed in \S\ref{sec:testbed}.
\noindent\textbf{Testbed Setup.} We perform testbed data collection in 2 homes, 1 office, and 2 open workspaces. Each home is in the standard apartment style including kitchen, living room, dining room, and bedrooms etc. The home settings allow user to perform any of the six scenarios, while the office and open workspaces are dominated by working, leisure, and recreation activities. Each participant is instructed to perform their daily routine normally rather than following pre-defined scripts. This allows us to collect and evaluate HOIs at a more natural density across the daily life. We show the distribution across participants, scenarios, and locations in Tab.~\ref{tab:distribution}. We show the distribution of activities on time horizon in Fig.~\ref{fig:testbed-time-of-day} and example actions and objects in Fig.~\ref{fig:testbed-wordcloud}. 

\begin{table}[t]
\centering
\caption{Recording hours and participants (P1--10) of the real-world testbed per location and scenario.}
\vspace{-0.1in}
\label{tab:testbed-location}
\setlength{\tabcolsep}{1pt}
\newcommand{\twoline}[1]{\begin{tabular}[c]{@{}c@{}}#1\end{tabular}}
\resizebox{\columnwidth}{!}{%
\begin{tabular}{l*{6}{c}c}
\toprule
\textbf{Location} & \textbf{Leisure} & \textbf{Working} & \textbf{Eating} & \textbf{Cooking} & \textbf{Housework} & \textbf{Recreation} & \textbf{Total} \\
\midrule
Home1 & \twoline{2.5\\(P1, P2, P6)} & \twoline{2.6\\(P1, P2)} & \twoline{0.8\\(P1, P2, P6)} & \twoline{1.6\\(P1, P6)} & \twoline{1.6\\(P1, P2, P6)} & \twoline{2.2\\(P1, P2, P6)} & 11.3 \\
Home2 & -- & \twoline{3.4\\(P4)} & \twoline{0.3\\(P4)} & \twoline{0.8\\(P4)} & -- & -- & 4.5 \\
Workspace1 & \twoline{0.3\\(P5)} & \twoline{0.5\\(P3)} & \twoline{0.5\\(P1)} & -- & -- & -- & 1.3 \\
Workspace2 & \twoline{0.3\\(P9)} & \twoline{0.7\\(P3, P10)} & -- & -- & -- & \twoline{0.8\\(P8, P9)} & 1.7 \\
Office & \twoline{0.4\\(P7)} & \twoline{0.7\\(P10)} & -- & -- & -- & -- & 1.1 \\
\midrule
Total & 3.5 & 7.8 & 1.6 & 2.4 & 1.6 & 2.9 & 19.8 \\
\bottomrule
\end{tabular}}
\vspace{-0.1in}
\label{tab:distribution}
\end{table}

\clearpage
\bibliographystyle{unsrt}
\bibliography{references}

\end{document}